\documentclass[10pt,american,aps,prd,superscriptaddress,nofootinbib,notitlepage,preprintnumbers]{revtex4-2}
\usepackage[T1]{fontenc}
\usepackage[latin9]{inputenc}
\usepackage{amsmath}

\makeatletter
\def\Mpl{M_{\rm P}}
\usepackage{tikz}

\usepackage{babel}

\makeatother

\usepackage{babel}
\begin{document}
\preprint{YITP-22-39, IPMU22-0020}
\title{VCDM and Cuscuton}
\author{Antonio De Felice}
\email{antonio.defelice@yukawa.kyoto-u.ac.jp}

\affiliation{Center for Gravitational Physics and Quantum Information, Yukawa Institute
for Theoretical Physics, Kyoto University, 606-8502, Kyoto, Japan}
\author{Kei-ichi Maeda}
\email{maeda@waseda.jp}
\affiliation{Department of Physics, Waseda University, Shinjuku, Tokyo 169-8555,
  Japan}
\affiliation{Center for Gravitational Physics and Quantum Information, Yukawa Institute
for Theoretical Physics, Kyoto University, 606-8502, Kyoto, Japan}
\author{Shinji Mukohyama}
\email{shinji.mukohyama@yukawa.kyoto-u.ac.jp}

\affiliation{Center for Gravitational Physics and Quantum Information, Yukawa Institute
for Theoretical Physics, Kyoto University, 606-8502, Kyoto, Japan}
\affiliation{Kavli Institute for the Physics and Mathematics of the Universe (WPI),
The University of Tokyo, Kashiwa, Chiba 277-8583, Japan}
\author{Masroor C.\ Pookkillath}
\email{masroor.cp@yukawa.kyoto-u.ac.jp}

\affiliation{Center for Gravitational Physics and Quantum Information, Yukawa Institute
for Theoretical Physics, Kyoto University, 606-8502, Kyoto, Japan}
\date{\today}
\begin{abstract}
We investigate two Type-IIa Minimally Modified Gravity theories, namely
VCDM and Cuscuton theories. We confirm that all acceptable Cuscuton
solutions are always solutions for VCDM theory. However, the inverse
does not hold. We find that VCDM allows for the existence of exact
General Relativity (GR) solutions with or without the presence of
matter fields and a cosmological constant. We determine the conditions
of existence for such GR-VCDM solutions in terms of the trace of the
extrinsic curvature and on the fields which define the VCDM theory.
On the other hand, for the Cuscuton theory, we find that the same
set of exact GR solutions (such as Schwarzschild and Kerr spacetimes)
is not compatible with timelike configurations of the Cuscuton field
and therefore cannot be considered as acceptable solutions. Nonetheless,
in Cuscuton theory, there could exist solutions which are not the
same but close enough to GR solutions. We also show the conditions
to determine intrinsic-VCDM solutions, i.e.\ solutions which differ
from GR and do not belong to the Cuscuton model. We finally show that
in cosmology a mapping between VCDM and the Cuscuton is possible,
for a generic form of the VCDM potential. In particular, we find that
for a quadratic potential in VCDM theory, this mapping is well defined
giving an effective redefinition of the Planck mass for the cosmological
background solutions of both theories. 
\end{abstract}
\maketitle

\section{Introduction}


Even though General Relativity (GR) is a successful theory of gravity,
it still needs to explain the dark sector of our universe at large
scales in a way the theory and experiments/observations can agree
with each other. Hence, exploring modified gravity theories at the
cosmological scales has been showing a constantly growing interest~\citep{Clifton:2011jh,Tsujikawa:2010zza}.
In most cases, modified gravity theories introduce some additional
degrees of freedom, which are not present in GR. For example, in the
scalar-tensor theories of gravity, in addition to the two polarizations
of the gravitational waves, we typically have an additional propagating
scalar mode~\citep{Fujii:2003pa}. Whereas in vector-tensor theories
of gravity, one expects to find five propagating degrees of freedom,
in general~\citep{Heisenberg:2017mzp,DeFelice:2016yws,DeFelice:2020sdq}
(in addition to the standard model fields). Since all the modifications
are amending the Einstein-Hilbert action, it is natural to study the
existence and validity of solutions of these modified gravity models
also beyond cosmology, describing e.g.\ other gravitational systems
like black-holes, stars, etc. To pass the astrophysical constraints
for these new theories, one typically needs some kind of screening
mechanisms at least at solar system scales to hide the otherwise additional
propagating modes~\citep{Joyce:2014kja,Brax:2015cla,Koyama:2015vza}.

On the other hand, there has been a recent development in a class
of modified gravity theories, generally called Minimally Modified
Gravity (MMG)~\citep{DeFelice:2015hla,Lin:2017oow,Mukohyama:2019unx,DeFelice:2020eju,Aoki:2020lig}.
These theories do not contain any additional local degrees of freedom other
than those that are present in GR. This minimalist's
approach is aimed at avoiding the constraints connected to the existence
of extra degrees of freedom. The MMG theories are classified into
Type-I and Type-II, where Type-I theories are endowed with an Einstein
frame and Type-II not~\citep{Aoki:2018brq}. If an MMG has the same
propagation speed for both electromagnetic and gravitational waves,
then this model is classified as Type-Ia or Type-IIa. On the other
hand, if the propagation speed is different between electromagnetic
waves and gravitational waves, it is classified as Type-Ib or Type-IIb~\citep{Aoki:2021zuy}.
Several investigations have been performed for these theories both
in the context of astrophysics and cosmology~\citep{DeFelice:2015moy,DeFelice:2018vza,Aoki:2020oqc,DeFelice:2021trp,Aoki:2020iwm,Aoki:2020ila,deAraujo:2021cnd,Pookkillath:2021gdp}.

One example of such MMG theory was introduced very recently~\cite{DeFelice:2020eju}.
It is a Type-IIa theory~\citep{Aoki:2021zuy} and it is dubbed VCDM theory\footnote{This theory should not be confused with other ``VCDM'' theories,
such as those introduced in \cite{Parker:1999td,BeltranJimenez:2008iye}.}. The construction of the VCDM theory is the following: 1) perform
a canonical transformation of GR Hamiltonian; 2) add a cosmological
constant in the new canonical frame; 3) add a gauge fixing term which
works as a constraint as to have only two degrees of freedom in the
gravity sector; 4) perform an inverse canonical transformation as
to have a resulting Hamiltonian which differs from GR; 5) make a Legendre
transformation in order to obtain the VCDM Lagrangian; 6) add standard
matter fields.

Some exact solutions of VCDM theory have been found and studied. In
particular, black-holes/vacuum solutions have been explored, see e.g.\
\cite{DeFelice:2020onz}. Although the theory, by construction, does
not possess any extra degrees of freedom, still it breaks the Birkhoff
theorem, and one needs to find the most general solutions compatible
with some symmetry and set the free parameters of the solutions either
by imposing appropriate boundary conditions, or by matching with observations.
This is due to the presence of a shadowy mode, which leads to the
presence of additional free parameters other than mass and the cosmological
constant. The spherically symmetric static star solutions were also
studied in the context of the VCDM theory\footnote{The same spherically symmetric static star solution valid in VCDM
is also valid for another Type-II MMG theory named VCCDM~\citep{DeFelice:2020prd}.}. It was shown that once we fix the physical boundary for the Misner-Sharp
mass of the system the solution exactly matches those of GR~\citep{DeFelice:2021xps}.
The cosmology of the VCDM theory was also explored and it was shown
that the $H_{0}$ tension can be reduced/addressed within this theory~\citep{DeFelice:2020cpt},
since the theory allows for general dynamics for $H(z)$ (with $H(z)>0$)
without introducing unstable/ghost degrees of freedom.

Another Type-IIa theory that is discussed in the literature is the
Cuscuton theory~\citep{Afshordi:2006ad}. If one starts from a scalar
tensor theory which, to the standard Einstein Hilbert term, adds a
term in the form $P(X,\varphi)=\mu^{2}\sqrt{-X}-U(\varphi)$, where
$X=(\partial\varphi)^{2}$, provided that the scalar field $\varphi$
is timelike (and this proves to be a crucial assumption), then in
the unitary gauge ($\varphi=t$), it is straightforward to show that
the theory has only two degrees of freedom coming from the gravity
sector. This theory, for a timelike field $\varphi$, defines the
Cuscuton theory, which can be regarded, a posteriori, as being a Type-IIa
MMG theory~\citep{Aoki:2021zuy}. Many aspects of the Cuscuton theory
have already been explored, see e.g.\ the following references~\citep{Afshordi:2007yx,Afshordi:2009tt,Boruah:2017tvg,Boruah:2018pvq,Iyonaga:2018vnu,Quintin:2019orx,Iyonaga:2020bmm,Panpanich:2021lsd,Maeda:2022ozc,Bartolo:2021wpt}.

In both these theories, VCDM and Cuscuton, there exists a scalar field
which is not propagating, leaving only two gravitational degrees of
freedom in the gravity sector. This scalar field is associated to
the existence of a shadowy mode, which, by definition, obeys an elliptic
equation of motion instead of a hyperbolic one. As mentioned above,
at least for known solutions within the VCDM theory, the influence
of the shadowy mode on background solutions can be removed if an appropriate
physical boundary condition is imposed. In other words, the behavior
of the shadowy mode is controlled by the physical boundary conditions
provided by the environment.

Since both these theories are Type-IIa MMG theories, it is interesting
to explore the differences between these two theories. To address
this question, it is a good idea to study the known non-perturbative
solutions allowed for these theories. On top of that, it is natural
to ask if the allowed solutions for both these theories can coincide
with solutions of GR or not, and if not, explore their difference.
In this work, we address all these issues. We find that under certain
conditions there exists a set of solutions in the VCDM which can be
exactly matched with those of GR. On the other hand, not all solutions
of VCDM are also GR solutions or Cuscuton solutions. Instead, the
solutions in the Cuscuton theory cannot exactly coincide with the
ones of GR otherwise the Cuscuton field would stop being timelike.
This last property of the Cuscuton field does not necessarily exclude
the phenomenology of this theory, since, after all, the solutions
do not need to be exactly equal to the ones of GR but only close enough
to them, compatibly with known experimental and observational constraints.

\begin{figure}[ht]
\begin{tikzpicture}
\draw[thick] (0,0) -- (7,0) -- (7,6.5) -- (0,6.5) -- (0,0);
\draw (3.5,3.25) circle (3cm);
\draw[line width=1.5mm, white] (3.5,0.2) -- (3.5,6.3);
\draw[thick] (3.45,0.25) -- (3.45,6.25);
\draw[dashed, thick] (3.55,0.25) -- (3.55,6.25);
\filldraw[black] (3.7,3.5)   node[anchor=west]{$\nabla^2\lambda_2=0,$};
\filldraw[black] (3.7,2.75)   node[anchor=west]{$V_{,\phi\phi}\neq 0$};
\filldraw[black] (3.7,4.25)   node[anchor=west]{\underline{Cuscuton}};
\filldraw[black] (5.5,6)   node[anchor=west]{\underline{VCDM}};
\filldraw[black] (1.7,3.5)   node[anchor=west]{$K=K_0,$};
\filldraw[black] (1.7,2.75)   node[anchor=west]{$\phi=\phi_{0},$};
\filldraw[black] (1.7,2)   node[anchor=west]{$V_{,\phi\phi}\neq 0$};
\filldraw[black] (0.7,4.25)   node[anchor=west]{\underline{GR with $K=K_0$}};
\end{tikzpicture} \caption{Summary of the results. We classify solutions (with or without matter)
for VCDM and the Cuscuton model. The dashed line corresponds to the
solutions in Cuscuton theory that are close enough to (but not exactly
equal to) the GR solutions. In the present paper we mainly focus on
the cases with $V_{,\phi\phi}\protect\ne0$ since the VCDM cosmology
with a linear potential is indistinguishable from the standard $\Lambda$CDM
both at the background and at the linear perturbation level and thus
is less motivated.}
\label{summaryFig} 
\end{figure}
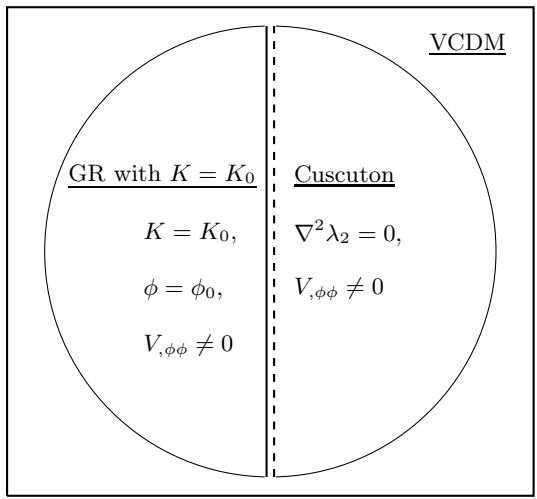

The figure~\ref{summaryFig} summarizes the results of this paper.
There exists a set of vacuum solutions of VCDM (with generic potential
$V$, i.e.\ satisfying $V_{,\phi\phi}\neq0$) which are also solutions
of GR (i.e.\ GR solutions in the presence of minimally coupled matter
fields and a cosmological constant) once we impose that both the extrinsic
curvature $K$ and the field $\phi$ to be constants (in space and
time). The left semicircle in the figure~\ref{summaryFig} represents
GR solutions in VCDM theory.

In the unitary gauge\footnote{Since the Cuscuton field is bound to be timelike, it is always possible
for acceptable solutions in Cuscuton theory to pick up the $\varphi$-unitary
gauge.}, solutions of the Cuscuton Lagrangian are also solutions of the VCDM
theory when we impose $\nabla^{2}\lambda_{2}=0$ ($\lambda_{2}$ can
be interpreted as being the shadowy mode in the VCDM theory), provided
that $V_{,\phi\phi}\neq0$ and $\varphi$ remains timelike, as first
shown in~\cite{Aoki:2021zuy}. As we will see later on, GR solutions
cannot be exact solutions of the Cuscuton theory (being $\varphi$
forced to remain timelike), however there are cases (at least known
examples in cosmology exist) for which Cuscuton solutions may be close
to GR, provided a well-behaved limit $(\partial\varphi)^{2}\to0$
exists. The dashed line in Figure~\ref{summaryFig} shows the Cuscuton
solutions that are close enough to (but not exactly equal to) GR.
Finally, solutions for which $\nabla^{2}\lambda_{2}\neq0$, and at
the same time $K$ is not a constant (in space or time) are VCDM-intrinsic
solutions, i.e.\ solutions which differ from GR and which do not
belong to the Cuscuton theory.

The rest of this paper is organized as follows. In section \ref{sec:GR-vacuum-solutions-in}
we investigate the condition under which the VCDM admits solutions
of $\Lambda$GR, i.e.\ solutions of GR in the presence of a cosmological
constant. We find these conditions by comparing the VCDM Hamilton
equations of motion for a general background to those of $\Lambda$GR.
We show in particular in this section that e.g.\ the Schwarzschild-de
Sitter and the Kerr-de Sitter are valid vacuum solutions in VCDM theory.
Then in section \ref{sec:vcdm_matter} we investigate VCDM solutions
with matter. In particular we study the weak field limit of this theory
and confirm that it reduces to that of GR, compatibly with previous
studies \citep{DeFelice:2020onz}. We also show that VCDM admits solutions
of GR in the presence of minimally coupled matter fields. For this
purpose we introduce a four dimensional covariant action which reduces
to the one of VCDM after choosing the unitary gauge for one of the
fields. Furthermore, we discuss the cosmological not-necessarily-flat
background and show the reconstruction of a given $H(z)$ for VCDM.
Subsequently, in section \ref{sec:vcdm_and_cuscuton} we briefly discuss
Cuscuton theory and discuss various backgrounds (including Schwarzschild
and Kerr ones) which are perfectly valid in VCDM, but which are, on
the other hand, not acceptable in the Cuscuton theory. We also investigate
an exact mapping on a cosmological background between VCDM and the
Cuscuton theory. Finally we give our concluding remarks in section
\ref{sec:Conclusions-and-discussion}.

\textbf{Notation:} the Latin letters are used for the three dimensional
spatial indices for example $a,b,c,\dots=1,2,3$, while the Greek
letter are used to indicate four dimensional spacetime indices $\alpha,\beta,\gamma,\dots=0,1,2,3$.
We work in the units for which $c=1$. Also we have the space time
metric signature convention $(-,+,+,+)$. Finally, by $V(\phi)$ we
will denote the potential term of VCDM theory and $\phi$ denotes
the scalar field of VCDM theory. Instead $U(\varphi)$ denotes the
potential in the Cuscuton theory and $\varphi$ denotes the Cuscuton
scalar field.


\section{Vacuum GR solutions in VCDM \label{sec:GR-vacuum-solutions-in}}


From the previous investigations of different spherically symmetric
solutions of VCDM theory (see e.g.\
\citep{DeFelice:2020onz,DeFelice:2021xps}), we know that there exist
solutions inside VCDM which are the same as those of GR, provided
that we set appropriate physical boundary conditions for the shadowy
mode, e.g., \ the finiteness of the (generalized) Misner-Sharp mass
for a spherically symmetric isolated compact gravitational body/system.
Nevertheless, even though these GR/VCDM solutions do exist, VCDM theory,
by construction, is different from GR. In particular, the presence
of the shadowy mode, by construction, implies the existence of a mode
whose spatial dependence is determined by an elliptic equation of
motion, which requires a preferred slicing where to set boundary conditions.
Therefore, by construction, the theory, since it requires fixing boundary
conditions on this field, is bound to pick up a natural slicing for
the theory which on the other hand breaks the general 4-D diffeomorphism
invariance. At the same time, the constraints which define the theory
are such that VCDM has the same number of gravitational propagating
modes of GR, namely the two standard tensor polarization of GR. In
summary VCDM differs from GR although it shares the same physical
degrees of freedom. Hence, it is natural to ask whether there exist
(or not) VCDM background solutions which exactly match GR solutions,
and if so, to determine the conditions of existence of such solutions.
In this section, we study these conditions of equality of the solutions
in VCDM theory compared to GR solutions in the presence, at most,
of a cosmological constant. We will extend this discussion in the
presence of matter fields in section \ref{subsec:matter-solutions}.


\subsection{Vacuum VCDM equations of motion}


In the following we will work in the VCDM-natural slicing, the one
which sets the shadowy mode to fulfill a Laplacian equation of motion.
After having chosen this slicing, we will make use of the standard
ADM splitting for the metric. Since we are looking for solutions which
are required, by assumption, to reduce to the same solutions of GR
for a generic background/slicing, then we have that the VCDM three
dimensional metric $\gamma_{ab}$, lapse $N$, and shift $N_{a}$
fields are the same as those of GR.

As a consequence, after finding the equations of motion for VCDM in
the unitary gauge for a generic background, we will consider the variables
$\gamma_{ab}$, $N$, and $N_{a}$ as satisfying also the equations
of motion of GR. This, in turn, will lead to imposing some nontrivial
conditions in the VCDM theory, that we want to determine.

On using the 1+3 ADM splitting for a generic background, we find it
convenient to determine the equations of motion by using the Hamiltonian
approach. Hence, first of all, we write down the Hamiltonian for VCDM
in vacuum as follows
\begin{eqnarray}
H & = & \int dtd^{3}x\sqrt{\gamma}\left\{ \lambda_{C}\left(\Mpl^{2}\phi-\gamma_{ab}\tilde{\pi}^{ab}\right)-2N^{a}D_{b}\tilde{\pi}_{a}{}^{b}-N\left[\frac{1}{2}\Mpl^{2}\,R-\frac{2}{\Mpl^{2}}\left(\tilde{\pi}^{ab}\tilde{\pi}_{ab}-\frac{1}{2}\tilde{\pi}^{a}{}_{a}\tilde{\pi}^{b}{}_{b}\right)-\Mpl^{2}V(\phi)\right]\right.\nonumber \\
 & + & \left.\Mpl^{2}\lambda_{{\rm gf}}^{a}D_{a}\phi+\lambda_{\phi}\tilde{\pi}_{\phi}\right\} ,\label{eq:vcdm_hamil}
\end{eqnarray}
whereas the presence of other matter fields will be discussed in section
\ref{subsec:matter-solutions}. In the expression of Eq.\ \eqref{eq:vcdm_hamil}
we have that $\lambda_{C}$, $\lambda_{\phi}$, $\lambda_{{\rm gf}}^{a}$,
$N^{a}$, $N$ are to be considered as Lagrange multipliers which
set all the constraints of the theory. Also, we have defined $\tilde{\pi}^{ab}\equiv\pi^{ab}/\sqrt{\gamma}$
and $\tilde{\pi}_{\phi}\equiv\pi_{\phi}/\sqrt{\gamma}$, where $\pi^{ab}$
and $\pi_{\phi}$ are the momenta conjugate to the metric variables
$\gamma_{ab}$ and the VCDM field $\phi$ respectively. Here and in
the following $R$ represents the 3-D Ricci scalar.

From the above VCDM Hamiltonian~(\ref{eq:vcdm_hamil}), for a generic
background, we have to fulfill all the following constraints, which
are set by the above mentioned Lagrangian multipliers: 
\begin{eqnarray}
0 & \approx & \mathcal{C}_{1}\equiv\sqrt{\gamma}\left[\frac{2}{\Mpl^{2}}\left(\tilde{\pi}^{ab}\tilde{\pi}_{ab}-\frac{1}{2}\tilde{\pi}^{a}{}_{a}\tilde{\pi}^{b}{}_{b}\right)-\frac{1}{2}\Mpl^{2}\,R+\Mpl^{2}V(\phi)\right]\,,\label{eq:C1_cons}\\
0 & \approx & \mathcal{C}_{2}\equiv\sqrt{\gamma}\,\tilde{\pi}_{\phi}\,,\label{eq:C2_cons}\\
0 & \approx & \mathcal{C}_{3}\equiv\sqrt{\gamma}\,(\Mpl^{2}\,\phi-\gamma_{ab}\tilde{\pi}^{ab})\,,\label{eq:C3_cons}\\
0 & \approx & \mathcal{C}_{4a}\equiv-2\sqrt{\gamma}D_{b}\tilde{\pi}_{a}{}^{b}+\sqrt{\gamma}\tilde{\pi}_{\phi}D_{a}\phi\,,\label{eq:C4_cons}\\
0 & \approx & \mathcal{C}_{5a}\equiv\sqrt{\gamma}\,\Mpl^{2}D_{a}\phi\,.\label{eq:C5_cons}
\end{eqnarray}
As a consequence of the constraints $\mathcal{C}_{3}$ and $\mathcal{C}_{5a}$,
we find that $D_{a}\tilde{\pi}^{b}{}_{b}\approx0$ on the surface
constraint. Now, setting the time evolution of these constraints to
vanish generically leads to equations which set the value for the
Lagrange multipliers on the solutions. However, for $\mathcal{C}_{4a}$,
this does not happen because they represent first class constraints
for the system\footnote{The term $\mathcal{C}_{4a}$ is the redefinition of the momentum constraint $\tilde{\mathcal{C}}_{4a}\equiv-2\sqrt{\gamma}D_{b}\tilde{\pi}_{a}{}^{b}\approx0$, which is given as $\mathcal{C}_{4a}\equiv\tilde{\mathcal{C}}_{4a}+\sqrt{\gamma}\tilde{\pi}_{\phi}D_{a}\phi\approx0$, and which is just a linear combination of constraints. However, with this redefinition the momentum constraint is now a first class constraint. In other words the Poisson bracket of $\mathcal{C}_{4a}$ with any other constraint vanishes. That is, there is a internal gauge freedom in the 3-D space, which ensures 3-D diffeomorphism invariance.} and they show that the three dimensional diffeomorphism invariance holds for this theory. In particular we find that 
\begin{equation}
\{\mathcal{C}_{2},H\}\approx0\,\to\qquad D_{a}\lambda_{{\rm gf}}^{a}=\lambda_{C}+N\,V_{,\phi}\,,
\end{equation}
where $\{f,g\}$ denotes the Poisson bracket of $f$ and $g$\footnote{More in detail $\{f,g\}=\sum_{i}\int d^{3}z\left(\frac{\delta f}{\delta q_{i}}\frac{\delta g}{\delta\pi^{i}}-\frac{\delta f}{\delta\pi^{i}}\frac{\delta g}{\delta q_{i}}\right)$, and the sum is over all the dynamical fields, $\gamma_{ab}$ and $\phi$ in this section.}. Also we have 
\begin{equation}
\{\mathcal{C}_{3},H\}\approx0\,\to\qquad D_{a}D^{a}N+N\left[V+\frac{1}{\Mpl^{4}}\left(\tilde{\pi}^{a}{}_{a}\tilde{\pi}^{b}{}_{b}-4\tilde{\pi}^{ab}\tilde{\pi}_{ab}\right)\right]+\lambda_{\phi}=0\,,\label{eq:C3_dot}
\end{equation}
which can be used to fix the lapse $N$. Then 
\begin{equation}
\{\mathcal{C}_{5a},H\}\approx0:\qquad D_{a}\lambda_{\phi}=0\,,
\end{equation}
which sets the field $\lambda_{\phi}$. The equation of motion which
instead fixes $\lambda_{C}$ is found as follows 
\begin{equation}
\{\mathcal{C}_{1},H\}\approx0:\qquad D_{a}D^{a}\lambda_{C}+\lambda_{C}\left[V+\frac{1}{\Mpl^{4}}\left(\tilde{\pi}^{a}{}_{a}\tilde{\pi}^{b}{}_{b}-4\tilde{\pi}^{ab}\tilde{\pi}_{ab}\right)\right]-\lambda_{\phi}V_{,\phi}\approx0\,.
\end{equation}
So far, the treatment was fully general. We can now proceed to find
the general dynamical equations of motion for VCDM. Let us start by
writing the following ones 
\begin{eqnarray}
\dot{\gamma}_{ab} & = & \{\gamma_{ab},H\}=\frac{2N}{\Mpl^{2}}\left(2\tilde{\pi}_{ab}-\gamma_{ab}\tilde{\pi}^{c}{}_{c}\right)+D_{a}N_{b}+D_{b}N_{a}-\lambda_{C}\gamma_{ab}\,,\label{eq:gamma_dot}\\
\dot{\phi} & = & \{\phi,H\}=\lambda_{\phi}\,,\label{eq:phi_dot}\\
\dot{\pi}_{\phi} & = & \{\pi_{\phi},H\}=0\,.\label{eq:piphi_dot}
\end{eqnarray}
From the equation (\ref{eq:gamma_dot}) we can find 
\begin{equation}
\tilde{\pi}^{ab}=\frac{\Mpl^{2}}{2}\,(K^{ab}-K\,\gamma^{ab})-\frac{\Mpl^{2}}{2}\,\frac{\lambda_{C}}{N}\,\gamma^{ab}\,,\label{eq:piab-1}
\end{equation}
where we have used the definition of the extrinsic curvature 
\begin{equation}
K_{ab}\equiv\frac{1}{2N}\,(\dot{\gamma}_{ab}-D_{a}N_{b}-D_{b}N_{b})\,,
\end{equation}
which can be used anywhere in the equations of motion as to write
them for the variable $K^{ab}$. Now we want to write the following
dynamical equations 
\begin{eqnarray}
\dot{\tilde{\pi}} & = & \{\tilde{\pi}^{a}{}_{a},H\}\approx-\Mpl^{2}N\left[V+\frac{1}{\Mpl^{4}}\left(\tilde{\pi}^{a}{}_{a}\tilde{\pi}^{b}{}_{b}-4\tilde{\pi}^{ab}\tilde{\pi}_{ab}\right)\right]-\Mpl^{2}D_{a}D^{a}N\nonumber \\
 & = & \Mpl^{2}\lambda_{\phi}=\Mpl^{2}\,\dot{\phi}\,,
\end{eqnarray}
where $\tilde{\pi}\equiv\tilde{\pi}^{a}{}_{a}$ and we have used the
constraint for $N$ Eq.\ (\ref{eq:C3_dot}) as well as $D_{a}\mathcal{C}_{3}=0$
from Eq.\ (\ref{eq:C3_cons}), (\ref{eq:C5_cons}) and (\ref{eq:piphi_dot}).

Also we find, by taking the trace of Eq.\ (\ref{eq:piab-1}), that
\begin{equation}
\tilde{\pi}=-\Mpl^{2}\,K-\frac{3}{2}\Mpl^{2}\,\frac{\lambda_{C}}{N}\,,\label{eq:pitrace_vcdm}
\end{equation}
which, on using also Eq.\ (\ref{eq:C3_cons}), reduces to 
\begin{equation}
\frac{\lambda_{C}}{N}=-\frac{2}{3}\,(K+\phi)\,.\label{eq:lambdaC_K}
\end{equation}
On replacing the above equation back into Eq.\ (\ref{eq:piab-1}),
we get 
\begin{eqnarray}
\tilde{\pi}^{ab} & = & \frac{\Mpl^{2}}{2}\,(K^{ab}-K\,\gamma^{ab})+\frac{1}{3}\,\Mpl^{2}\,(K+\phi)\,\gamma^{ab}\,.\label{eq:piab}
\end{eqnarray}
Then we write the full Hamilton equations for $\dot{\tilde{\pi}}^{ab}=\{\tilde{\pi}^{ab},H\}$,
as 
\begin{align}
\dot{\tilde{\pi}}^{ab} & =-\tfrac{1}{2}\Mpl^{2}NR^{ab}+\tfrac{1}{4}\Mpl^{2}\ \gamma^{ab}NR-\tfrac{1}{2}\Mpl^{2}\gamma^{ab}NV+\tfrac{5}{2}\lambda_{C}\tilde{\pi}^{ab}-\frac{4N\tilde{\pi}^{ac}\tilde{\pi}^{b}{}_{c}}{\Mpl^{2}}+\frac{3N\tilde{\pi}^{ab}\tilde{\pi}^{c}{}_{c}}{\Mpl^{2}}+\frac{\gamma^{ab}\ N\tilde{\pi}_{cd}\tilde{\pi}^{cd}}{\Mpl^{2}}\nonumber \\
 & -\frac{\gamma^{ab}\ N\tilde{\pi}^{c}{}_{c}\tilde{\pi}^{d}{}_{d}}{2\Mpl^{2}}-\tfrac{1}{2}\Mpl^{2}\gamma^{ab}\lambda_{C}\phi+\tfrac{1}{2}\ \Mpl^{2}D^{b}D^{a}N+N^{c}D_{c}\tilde{\pi}^{ab}-\tfrac{1}{2}\ \Mpl^{2}\gamma^{ab}D_{c}D^{c}N-\tilde{\pi}^{b}{}_{c}D^{c}N^{a}\ -\tilde{\pi}^{a}{}_{c}D^{c}N^{b}\,,
\end{align}
where the left hand side can be replaced with the time derivative
of Eq.\ (\ref{eq:piab}). So far, we have considered the Hamilton
equations of motion for a general background in VCDM. We can now proceed
to find the conditions for them to be satisfied also by $\Lambda$GR
solutions.


\subsection{Solutions in VCDM which reduce to $\Lambda$GR solutions}


Let us now consider $\Lambda$GR solutions. This means we consider
the same solution for the lapse $N$, the shift $N_{a}$, and the
spatial 3-D metric $\gamma_{ab}$ which exist for GR on a given slicing.
This also implies that the expressions of $K_{ab}$ in VCDM and $\Lambda$GR
will coincide on this slicing. It should be noticed that, in the case
of $\Lambda$GR, we have only first class constraints, so that all
the Lagrange multipliers $N$ and $N^{a}$ cannot be determined by
the Hamiltonian procedure. By calling the GR-momentum conjugate to
$\gamma_{ab}$ as $\Pi^{ab}$, then in $\Lambda$GR, we find that
the equations of motion $\dot{\gamma}_{ab}=\{\gamma_{ab},H\}$ lead
to
\begin{equation}
\tilde{\Pi}^{ab}=\frac{\Mpl^{2}}{2}\,(K^{ab}-K\,\gamma^{ab})\,,\label{eq:piGR2K}
\end{equation}
where we have $\tilde{\Pi}^{ab}\equiv\Pi^{ab}/\sqrt{\gamma}$. Taking
the trace of the above relation we also get 
\begin{equation}
\tilde{\Pi}\equiv\tilde{\Pi}^{a}{}_{a}=-\Mpl^{2}\,K\,.\label{eq:pigr2K}
\end{equation}
The $\Lambda$GR constraints can be written as 
\begin{eqnarray}
\frac{\mathcal{C}_{1}^{{\rm GR}}}{\sqrt{\gamma}} & = & \Mpl^{2}\Lambda-\tfrac{1}{2}\Mpl^{2}\ R+\frac{2(\tilde{\Pi}_{ab}\tilde{\Pi}^{ab}-\tfrac{1}{2}\tilde{\Pi}^{a}{}_{a}\tilde{\Pi}^{b}{}_{b})}{\Mpl^{2}}\approx0\,,\label{eq:GR_hamilt}\\
\frac{\mathcal{C}_{4a}^{{\rm GR}}}{\sqrt{\gamma}} & = & -2D_{b}\tilde{\Pi}_{a}{}^{b}\approx0\,.\label{eq:GR_moment}
\end{eqnarray}
On using the Hamilton equations of motion, we can also find the time
evolution of conjugate momenta. For instance, we have
\begin{align}
\dot{\tilde{\Pi}} & =-\tfrac{3}{2}\Mpl^{2}\Lambda N+\tfrac{1}{4}\Mpl^{2}NR+\ \frac{3N\tilde{\Pi}_{ab}\tilde{\Pi}^{ab}}{\Mpl^{2}}-\frac{N\tilde{\Pi}^{a}{}_{a}\tilde{\Pi}^{b}{}_{b}}{2\Mpl^{2}}+N^{a}D_{a}\ \tilde{\Pi}^{b}{}_{b}-\Mpl^{2}D_{a}D^{a}N\nonumber \\
 & \approx-\Mpl^{2}\Lambda N+\frac{4N\tilde{\Pi}_{ab}\tilde{\Pi}^{ab}}{\Mpl^{2}}-\frac{N\tilde{\Pi}^{a}{}_{a}\tilde{\Pi}^{b}{}_{b}}{\Mpl^{2}}+N^{a}D_{a}\ \tilde{\Pi}^{b}{}_{b}-\Mpl^{2}D_{a}D^{a}N\,,\label{eq:TracePiGR}
\end{align}
where we have used the Hamiltonian constraint, Eq.\ (\ref{eq:GR_hamilt}).
Furthermore we have that 
\begin{align}
\dot{\tilde{\Pi}}^{ab} & =-\tfrac{1}{2}\Mpl^{2}\Lambda\gamma^{ab}N-\tfrac{1}{2}\ \Mpl^{2}NR^{ab}+\tfrac{1}{4}\Mpl^{2}\gamma^{ab}NR-\frac{4N\tilde{\Pi}^{ac}\tilde{\Pi}^{b}{}_{c}}{\Mpl^{2}}+\frac{3N\tilde{\Pi}^{ab}\tilde{\Pi}^{c}{}_{c}}{\Mpl^{2}}+\frac{\gamma^{ab}N\tilde{\Pi}_{cd}\tilde{\Pi}^{cd}}{\Mpl^{2}}\nonumber \\
 & -\frac{\ \gamma^{ab}N\tilde{\Pi}^{c}{}_{c}\tilde{\Pi}^{d}{}_{d}}{2\ \Mpl^{2}}+\tfrac{1}{2}\Mpl^{2}D^{b}D^{a}N+N^{c}D_{c}\tilde{\Pi}^{ab}-\tfrac{1}{2}\Mpl^{2}\gamma^{ab}D_{c}D^{c}N-\tilde{\Pi}^{b}{}_{c}D^{c}N^{a}-\tilde{\Pi}^{a}{}_{c}D^{c}N^{b}\,.
\end{align}
We are now ready to study the conditions under which the solutions
of VCDM and GR coincide with each other, at least locally. Here the
logic is to apply the above GR-solutions as to constrain the VCDM
Hamiltonian constraints/equations of motion.

The first thing we notice by comparing Eq.\ (\ref{eq:piab-1}) with Eq.\ (\ref{eq:piGR2K}) is that
\begin{equation}
\tilde{\pi}^{ab}=\tilde{\Pi}^{ab}-\frac{\Mpl^{2}}{2}\,\frac{\lambda_{C}}{N}\,\gamma^{ab}\,.\label{eq:pi2piGR}
\end{equation}
On applying the operator $D_{b}$ on both sides of this equation and
on using the momentum constraint in VCDM Eq.\ (\ref{eq:C4_cons})
together with the gauge fixing constraint Eq.\ (\ref{eq:C5_cons})
and Eq.\ (\ref{eq:GR_moment}), we find 
\begin{equation}
D_{a}(\lambda_{C}/N)=0\,.\label{eq:condition-1}
\end{equation}

On using Eq.\ (\ref{eq:pi2piGR}) the Hamiltonian constraint in VCDM
written in terms of $\tilde{\Pi}^{ab}$ is 
\begin{equation}
\frac{\mathcal{C}_{1}}{\sqrt{\gamma}}=\Mpl^{2}\left[-\frac{3\lambda_{C}^{2}}{4N^{2}}-\tfrac{1}{2}R+V\right]+\frac{\lambda_{C}\tilde{\Pi}^{a}{}_{a}}{N}+\ \frac{2\tilde{\Pi}_{ab}\tilde{\Pi}^{ab}-\tilde{\Pi}^{a}{}_{a}\tilde{\Pi}^{b}{}_{b}}{\Mpl^{2}}=0\,.
\end{equation}
On comparing the above expression with the Hamiltonian constraint
Eq.\ (\ref{eq:GR_hamilt}) of $\Lambda$GR, and using Eq.\ (\ref{eq:pigr2K}),
we require a second condition to hold, namely 
\begin{equation}
\Lambda=V(\phi)-\frac{3}{4}\,\frac{\lambda_{C}^{2}}{N^{2}}-\frac{\lambda_{C}}{N}\,K\,.\label{eq:condition-2}
\end{equation}
By taking a spatial covariant derivative of the above expression,
we reach another condition, namely 
\begin{equation}
D_{a}K=0\,,\qquad{\rm or}\qquad D_{a}\tilde{\Pi}^{b}{}_{b}=0\,,\label{eq:condition-3}
\end{equation}
where we have used the gauge constraint Eq.\ (\ref{eq:C5_cons})
and Eq.\ (\ref{eq:condition-1}). For the special case $\lambda_{C}=0$,
from Eq.\ (\ref{eq:lambdaC_K}) we know that $K=-\phi$, which, after
taking a covariant derivative and using Eq.\ (\ref{eq:C5_cons}),
again leads to the condition Eq.\ (\ref{eq:condition-3}).

In VCDM the time derivative of $\tilde{\pi}$, using also Eq.\ (\ref{eq:pitrace_vcdm}),
leads to 
\begin{align}
\dot{\tilde{\pi}} & =\dot{\tilde{\Pi}}-\frac{3}{2}\Mpl^{2}\,\frac{d}{dt}\!\left(\frac{\lambda_{C}}{N}\right)\nonumber \\
 & =\frac{3\Mpl^{2}\lambda_{C}{}^{2}}{4N}-\Mpl^{2}NV(\phi)-\ \lambda_{C}\tilde{\Pi}^{a}{}_{a}+\frac{4N\tilde{\Pi}_{ab}\tilde{\Pi}^{ab}}{\Mpl^{2}}-\frac{N\tilde{\Pi}^{a}{}_{a}\tilde{\Pi}^{b}{}_{b}}{\Mpl^{2}}\nonumber \\
 & -\frac{3\Mpl^{2}N^{a}\ D_{a}\lambda_{C}{}}{2N}+\frac{3\Mpl^{2}\lambda_{C}{}N^{a}\ D_{a}N}{2N^{2}}+N^{a}D_{a}\tilde{\Pi}^{b}{}_{b}-\Mpl^{2}\ D_{a}D^{a}N\nonumber \\
 & =\frac{3\Mpl^{2}\lambda_{C}{}^{2}}{4N}-\Mpl^{2}NV(\phi)-\ \lambda_{C}{}\tilde{\Pi}^{a}{}_{a}+\dot{\tilde{\Pi}}+\Mpl^{2}\Lambda N,
\end{align}
where we have used Eqs.\ (\ref{eq:pi2piGR}), (\ref{eq:TracePiGR})
and (\ref{eq:condition-3}). Then this result together with Eq.\ (\ref{eq:pigr2K})
lead to 
\begin{equation}
-\Lambda-\frac{3}{2}\,\frac{1}{N}\,\frac{d}{dt}\!\left(\frac{\lambda_{C}}{N}\right)=\frac{\lambda_{C}}{N}K+\frac{3\lambda_{C}^{2}}{4N^{2}}-V(\phi)\,.\label{eq:Lambda_K_lambdaC_dot_V}
\end{equation}
Comparing the condition Eq.\ (\ref{eq:condition-2}) with the above,
for consistency we reach the condition 
\begin{equation}
\frac{d}{dt}\!\left(\frac{\lambda_{C}}{N}\right)=0\,,\qquad{\rm or}\qquad\lambda_{C}=\lambda_{0}\,N\,,\qquad\lambda_{0}={\rm constant}\,.\label{eq:condition-4}
\end{equation}
The above relation, used in Eq.\ (\ref{eq:lambdaC_K}), leads to
\begin{equation}
K=-\phi-\frac{3}{2}\,\lambda_{0}\,,
\end{equation}
which also gives $\dot{K}=-\dot{\phi}$.

Substituting this last relation for $K$ together with Eq.\ (\ref{eq:condition-4})
into Eq.\ (\ref{eq:Lambda_K_lambdaC_dot_V}) and taking a time derivative
we obtain

\begin{equation}
\dot{\phi}\,(\lambda_{0}+V_{,\phi})=0\,,\label{eq-for-phiv}
\end{equation}
which is solved in general only for a constant $\phi$, and we will
not consider the case of a special linear form for the potential in
detail, as giving trivial results in cosmology\footnote{In fact, for the case of a linear potential $V=\beta_{0}+\beta_{1}\phi$,
we would have, as a possible solution of (\ref{eq-for-phiv}), that
$\lambda_{0}=-\beta_{1}$. This would not set $\phi$ to be necessarily
constant, leaving $\phi=\phi(t)$, as well as $K=K(t)$. This is what
actually happens in cosmology, as a linear potential makes VCDM solutions
exactly reduce to $\Lambda$CDM, see e.g.\ \cite{DeFelice:2020eju}.
However, even for a linear potential $V$, there could be non-trivial,
non-GR, VCDM-solutions when $K$ becomes space-and-time dependent.}. Therefore for a general VCDM potential, we have 
\begin{equation}
\phi=\phi_{0}\,,\qquad K=K_{0}=-\phi_{0}-\frac{3}{2}\,\lambda_{0}\,,\qquad\Lambda=\lambda_{0}\phi_{0}+\frac{3}{4}\,\lambda_{0}^{2}+V(\phi_{0})\,,\label{eq:phi_K_const_Lambda}
\end{equation}
or in other words when both $\phi$ and $K$ are constants VCDM is
equivalent to $\Lambda$GR with an effective cosmological constant
$\Lambda$ given by the expression in Eq.\ (\ref{eq:phi_K_const_Lambda}).
On considering the other equation of motion for $\dot{\tilde{\pi}}^{ab}$,
we find
\begin{equation}
\tfrac{1}{2}\Mpl^{2}\Lambda\gamma^{ab}N-\tfrac{3}{8}\ \Mpl^{2}\lambda_{0}{}^{2}\gamma^{ab}N-\Mpl^{2}\lambda_{0}{}\ \gamma^{ab}NK-\tfrac{1}{2}\Mpl^{2}\gamma^{ab}N\ V(\phi)-\lambda_{0}{}\gamma^{ab}N\tilde{\Pi}^{c}{}_{c}-\tfrac{1}{2}\Mpl^{2}\lambda_{0}{}\gamma^{ab}N\phi=0\,,
\end{equation}
 which can be shown to lead to 
\begin{equation}
\tfrac{1}{2}\Mpl^{2}\gamma^{ab}N\,[\Lambda-V(\phi_{0})-\tfrac{1}{4}\lambda_{0}(3\lambda_{0}+4\phi_{0})]=0\,,
\end{equation}
which is automatically satisfied.

As for the other Lagrange multipliers of VCDM, we find that 
\begin{equation}
\lambda_{\phi}=\dot{\phi}=0\,,
\end{equation}
whereas the equation of motion defining $N$ is automatically satisfied
as well as the one defining $\lambda_{C}=\lambda_{0}\,N$. The only
leftover nontrivial equation of motion is then 
\begin{equation}
D_{a}\lambda_{{\rm gf}}^{a}=N\left[V_{,\phi_{0}}-\frac{2}{3}(K_{0}+\phi_{0})\right],
\end{equation}
which can be used in order to solve for $\lambda_{{\rm gf}}^{a}$.
It should be noticed that VCDM-$\Lambda$GR solutions do not necessarily
have a vanishing $D_{a}\lambda_{{\rm gf}}^{a}$. These VCDM solutions
were first found, for the particular case of a static, spherically
symmetric background, in \cite{DeFelice:2020onz}, which were shown
to correspond to the Schwarzschild-de Sitter solutions of GR in the
constant-$K$ slicing.

In summary, any $\Lambda$GR solution in a constant-$K$ slicing can
be embedded in the VCDM theory as a solution. We have all relevant
equations that determine the VCDM fields once a $\Lambda$GR solution
and a constant-$K$ slicing are specified.


\subsection{Example: Kerr-de Sitter solutions}


As a lemma based on the previous discussion, for the special case
of $K_{0}=0$, i.e. in the maximal slicing, we have an effective cosmological
constant given by $\Lambda=V(\phi_{0})-\frac{1}{3}\,\phi_{0}^{2}$.
Here we are assuming that any nontrivial cosmological time dependence
for $\phi$ can be set to be negligible at astrophysical scales. Beside
the aforementioned case of the Schwarzschild-de Sitter solutions of
$\Lambda$GR first found in \citep{DeFelice:2020onz}, we want to
add here as a nontrivial case, the Kerr-de Sitter solutions in Boyer--Lindquist
coordinates, which describe the empty spacetime around an axisymmetric
distribution of matter. We are now going to show that they are solutions
not only for $\Lambda$GR, but also for the VCDM theory. In fact,
one has that the three dimensional line element for this background
in this slicing can be written as 
\begin{equation}
ds_{(3)}^{2}=\frac{a^{2}z^{2}+r^{2}}{\Delta}\,dr^{2}+\frac{a^{2}z^{2}+r^{2}}{\left(1+\frac{\Lambda\,a^{2}z^{2}}{3}\right)\left(1-z^{2}\right)}\,dz^{2}+\left[\left(a^{2}+r^{2}\right)^{2}\left(1+\frac{\Lambda\,a^{2}z^{2}}{3}\right)-\Delta\,a^{2}\left(1-z^{2}\right)\right]\frac{\left(1-z^{2}\right)d\theta_{2}{}^{2}}{\left(a^{2}z^{2}+r^{2}\right)\left(1+\frac{\Lambda\,a^{2}}{3}\right)^{2}}\,,
\end{equation}
where $\Delta=\left(a^{2}+r^{2}\right)\left(1-\frac{\Lambda\,r^{2}}{3}\right)-2mr$,
$z=\cos\theta_{1}$, and $\theta_{2}$ is the angle which defines
the axis of symmetry. Here $a$ is standard Kerr spin parameter and
$m$ is the mass parameter. Then for the same background solution,
the lapse and shift vector can be written as 
\begin{eqnarray}
\frac{1}{N^{2}} & = & \frac{\left[3\left(z^{2}-1\right)a^{2}\Delta+\left(a^{2}+r^{2}\right)^{2}\left(\Lambda\,a^{2}z^{2}+3\right)\right]\left(\Lambda\,a^{2}+3\right)^{2}}{9\Delta\left(a^{2}z^{2}+r^{2}\right)\left(\Lambda\,a^{2}z^{2}+3\right)}\,,\\
N_{a}\,dx^{a} & = & \left[\Delta-\left(1+\frac{\Lambda\,a^{2}z^{2}}{3}\right)\left(a^{2}+r^{2}\right)\right]\frac{a\left(1-z^{2}\right)}{\left(a^{2}z^{2}+r^{2}\right)\left(1+\frac{\Lambda\,a^{2}}{3}\right)^{2}}\,d\theta_{2}\,,
\end{eqnarray}
which lead to 
\begin{equation}
K=\gamma^{ab}K_{ab}=\frac{\gamma^{ab}}{2N}\,(\dot{\gamma}_{ab}-D_{a}N_{b}-D_{b}N_{a})=-\frac{1}{N}\,\gamma^{ab}D_{(a}N_{b)}=0\,,
\end{equation}
and confirms that this GR-solution is also a solution for VCDM. We
also discuss the existence of the McVittie solution in VCDM theory
later on.


\section{VCDM solutions with matter}

\label{sec:vcdm_matter} 


\subsection{Weak field solutions}

\label{sub-sec:weak_filed} 

Let us consider the weak field limit, namely a situation in which
the matter fields are supposed to source small perturbations around
the Minkowski background. The 3D metric, the lapse and the shift can
be written as 
\begin{eqnarray}
ds_{3}^{2} & = & (1+2\zeta)\,\delta_{ij}\,dx^{i}\,dx^{j}\,.\\
N & = & 1+\alpha\,,\\
N_{i} & = & \partial_{i}\chi\,,
\end{eqnarray}
whereas the VCDM fields are instead given by 
\begin{eqnarray}
\phi & = & \phi(t)+\delta\phi\,,\\
\lambda_{{\rm gf}}^{i} & = & \delta^{ij}\partial_{j}\delta\lambda_{2}\,,\\
\lambda & = & \lambda(t)+\delta\lambda\,.
\end{eqnarray}
Here $\lambda$ is a Lagrangian multiplier related to the field $\lambda_{C}$
introduced in the VCDM Hamiltonian (\ref{eq:vcdm_hamil}) by $\lambda=\lambda_{C}/N$,
and $\delta\lambda_{2}$ corresponds to the perturbation of the shadowy
mode present in the VCDM theory as will be explained in the discussion
after Eq.\ (\ref{eq:shadow}).

For the Minkowski background, the VCDM equations of motion lead to
\begin{eqnarray}
\phi(t) & = & \phi_{0}\,,\\
V & = & \frac{1}{3}\,\phi_{0}^{3}\,,\\
V_{,\phi} & = & \frac{2}{3}\,\phi_{0}\,,\\
\lambda(t) & = & -\frac{2}{3}\,\phi_{0}\,.
\end{eqnarray}
These are compatible with our previous finding connecting GR solutions
to VCDM solutions. Looking for the first non trivial corrections,
we find the effective Einstein tensor elements and set them equal
to the stress-energy tensor elements of a fluid, whereas the equations
of motion in VCDM which are not sourced by the matter fields are then
solved by themselves. For example we have 
\begin{equation}
\nabla^{2}\delta\phi=0\,,
\end{equation}
where $\nabla^{2}=\delta^{ij}\partial_{i}\partial_{j}$ on this background.
Along the same lines at leading order (assuming no shear and the fluid
velocity to be nonzero, but of sub-leading order): 
\begin{eqnarray}
\nabla^{2}\chi & = & \delta\phi\,,\\
\delta\lambda & = & -2\dot{\zeta}\,,\\
\nabla^{2}\zeta & = & -\nabla^{2}\alpha-\nabla^{2}\dot{\chi}\,,\\
2\Mpl^{2}\nabla^{2}\zeta & = & -\rho\,,
\end{eqnarray}
with $\rho$ satisfying the standard continuity equation as expected.
We find that $\nabla^{4}\zeta=-\nabla^{4}\alpha$, which on imposing
appropriate boundary conditions at infinity, leads to the same results
of GR, namely $\zeta=-\alpha$, and the standard Poisson equation
for the Newtonian potential.

For the tensor mode, this theory does not modify the dispersion relation
from that of GR. Hence this theory is called as Type-IIa MMG theory~\citep{Aoki:2021zuy}.


\subsection{Covariant action and GR solutions with matter fields}

\label{subsec:matter-solutions} 

In this subsection we show that under a certain condition, a solution
of GR in the presence of a cosmological constant and minimally coupled
matter fields can be embedded in VCDM as a consistent solution. For
this purpose it is convenient to use a covariant theory which reduces
to VCDM in the unitary gauge for the time coordinate. In the following
$\phi$, $\alpha$, $T$, are 4D scalar fields, and their connection
with other geometrical objects is determined by the Lagrange multipliers
$\lambda$, $\lambda_{2}$ and $\lambda_{T}$. Let us start by writing
the following gravitational action 
\begin{eqnarray}
S_{g} & = & \Mpl^{2}\int d^{4}x\sqrt{-g}\left\{ \tfrac{1}{2}\,R^{(4)}-V(\phi)-\tfrac{3}{4}\,\lambda^{2}-\lambda\,(\nabla^{\sigma}n_{\sigma}+\phi)\right.\nonumber \\
 & + & \left.\frac{\lambda_{2}}{\alpha}\,[\gamma^{\tau\rho}\nabla_{\tau}\nabla_{\rho}\phi+n^{\rho}(\nabla_{\rho}\phi)\,\nabla^{\sigma}n_{\sigma}]+\lambda_{T}\,(1+g^{\mu\nu}n_{\mu}n_{\nu})\right\} ,\label{eq:cov_act}\\
n_{\mu} & \equiv & -\alpha\,\nabla_{\mu}T\,,\\
\gamma^{\mu\nu} & = & g^{\mu\nu}+n^{\mu}n^{\nu}\,.
\end{eqnarray}
After integrating out the field $\alpha$ by using the equation of
motion for $\lambda_{T}$, we find 
\begin{eqnarray}
S_{g} & = & \Mpl^{2}\int d^{4}x\sqrt{-g}\left\{ \tfrac{1}{2}\,R^{(4)}-V(\phi)-\tfrac{3}{4}\,\lambda^{2}-\lambda\,(\nabla^{\sigma}n_{\sigma}+\phi)\right.\nonumber \\
 & + & \left.(-g^{\mu\nu}\nabla_{\mu}T\,\nabla_{\nu}T)^{1/2}\,\lambda_{2}\,[\gamma^{\tau\rho}\nabla_{\tau}\nabla_{\rho}\phi+n^{\rho}(\nabla_{\rho}\phi)\,\nabla^{\sigma}n_{\sigma}]\right\} ,\label{eq:covariant_vcdm_action}\\
n_{\mu} & = & -(-g^{\mu\nu}\nabla_{\mu}T\,\nabla_{\nu}T)^{-1/2}\,\nabla_{\mu}T\,,\\
\gamma^{\mu\nu} & = & g^{\mu\nu}+n^{\mu}n^{\nu}\,,
\end{eqnarray}
so that $\nabla_{\mu}T$ is, by construction, timelike. Then on choosing
$T=t$, not as the solution of some equations of motion, but rather
as a free choice of the time coordinate, we find the following action
\begin{equation}
S_{g}=\int d^{4}xN\sqrt{\gamma}\left[\frac{\Mpl^{2}}{2}\,[R+K_{ij}K^{ij}-K^{2}-2V(\phi)]+\frac{1}{N}\,\lambda_{2}\,\Mpl^{2}\gamma^{ij}D_{i}D_{j}\phi-\frac{3\Mpl^{2}\lambda^{2}}{4}-\Mpl^{2}\,\lambda\,(K+\phi)\right].\label{eq:vcdm_lag_uni}
\end{equation}
This agrees with the action of VCDM. Notice that $\lambda_{2}$ imposes
an elliptic equation on $\phi$, and vice versa $\phi$ imposes a
Laplacian operator on $\lambda_{2}$. Therefore the original VCDM
action can be thought of being the action of Eq.\ (\ref{eq:covariant_vcdm_action})
written in $T$-unitary-gauge.

It should be noted that we can integrate out the field $\lambda$
by using its own equation of motion\footnote{Here we can integrate out the Lagrange multiplier $\lambda$ because
its equation of motion is purely algebraic, getting a Lagrangian equivalent
to the VCDM Lagrangian.} 
\begin{equation}
\lambda=-\frac{2}{3}\,(K+\phi)\,.
\end{equation}
Furthermore, for a generic potential $V$, we can also integrate out
the field $\phi$ by using its own algebraic equation of motion\footnote{We should avoid the temptation of integrating out $\phi$, by solving
the differential equation $D^{2}\phi=0$ imposed by the field $\lambda_{2}$
at the level of the Lagrangian, not being an algebraic equation. In
fact, this in general leads to a different theory. For instance, on
considering an analogue case, i.e.\ having a similar structure, in
classical mechanics, take the following simple model $\mathcal{L}=\lambda_{2}\dot{q}-m^{2}\,q^{2}$.
On integrating out $q$ by solving the differential equation imposed
by $\lambda_{2}$ as $q=q_{0}$ would lead to a nonequivalent Lagrangian
$\mathcal{L}=-m^{2}q_{0}^{2}$, which gives no more dynamics for any
variable. Instead, one should first integrate by parts $\dot{q}$
giving $\mathcal{L}=-q\dot{\lambda}_{2}-m^{2}q^{2}$, and then integrating
out $q$, which has become now a Lagrange multiplier, by using its
own algebraic equation of motion, $q=-\dot{\lambda}_{2}/(2m^{2})$,
leads to a reduced Lagrangian $\mathcal{L}=\dot{\lambda}_{2}^{2}/(4m^{2})$,
out of which one finds equivalent equations of motion.} 
\begin{equation}
\phi=F\bigl[(\gamma^{ij}D_{i}D_{j}\lambda_{2})/N+\tfrac{2}{3}K\bigr]\,.\label{eq:shadow}
\end{equation}
Finally the VCDM Lagrangian can be written only in terms of the metric
variables and $\gamma^{ij}D_{i}D_{j}\lambda_{2}$, the shadowy mode,
whose equation of motion is clearly elliptical. This latter field
cannot be further integrated out, unless we introduce non-local terms
into the action, avoiding in this way the Lovelock theorem. Although
we have found a covariant theory which reduces to VCDM, the choice
of the slicing $T=t$ is precisely chosen because of the presence
of the shadowy mode. In fact, the shadowy mode, by its own equation
of motion, sets a preferred frame on which its elliptic differential
operator is defined. Then the $T$-unitary gauge is the natural choice
for the time coordinate for the above VCDM-covariant Lagrangian. Although
this covariant action may seem a redundant knowledge, nonetheless,
in same cases, one can use it in a proficient way, for example when
the $T$-equation of motion is needed (which is written in Eq.\ (\ref{eq:T-eom})
of Appendix \ref{sec:Covariant-VCDM-Tmunu}, and which, in unitary
gauge, can be found only after an appropriate manipulation the other
equations of motion) or when it is helpful to have an explicit expression
for $T_{\mu\nu}$ (even when it is evaluated, after finding it, in
unitary gauge).

Let us now use the covariant action of VCDM, introduced in Eq.\ \eqref{eq:cov_act},
in order to show that VCDM indeed admits GR solutions with minimally
coupled matter. The modified Einstein equations in covariant VCDM
can be written as 
\begin{equation}
\Mpl^{2}\,G^{\mu}{}_{\nu}=T_{~\nu}^{\mu}+{\mathcal{T}_{{\bf v}}}_{~\nu}^{\mu}\,,
\end{equation}
where $T_{~\nu}^{\mu}$ stands for the total matter field stress energy
tensor (i.e.\ excluding the VCDM contribution). Let us try to find
the condition under which we can embed GR solutions in VCDM. In this
case we require that ${\mathcal{T}_{{\bf v}}}_{~\nu}^{\mu}$ should
give a cosmological constant contribution. Therefore, as we have also
seen in the vacuum case, let us consider the case of $\phi=\phi_{0}={\rm constant}$.
Furthermore, let us assume that the solution admits $K=K_{0}={\rm constant}$,
where $K=\nabla^{\sigma}n_{\sigma}$ is the trace of the extrinsic
curvature induced by the $T$-coordinate choice\footnote{This corresponds to a constant-$K$ slicing.}.
In this case, the equation of motion for $\lambda$ sets also $\lambda$
itself to be a constant, i.e.\ $\lambda=\lambda_{0}$, on this background,
independently of the presence of matter fields since 
\begin{equation}
\lambda=-\frac{2}{3}\,(\nabla^{\sigma}n_{\sigma}+\phi)=-\frac{2}{3}\,(K_{0}+\phi_{0})=\lambda_{0}\,.
\end{equation}
Now, the equation of motion for $\alpha$, corresponding to Eq.\ \eqref{eq:E_alpha}
of Appendix \ref{sec:Covariant-VCDM-Tmunu}, evaluated for a constant
$\lambda$ and $\phi$, sets the following constraint on the solution
\begin{equation}
\frac{2\Mpl^{2}\lambda_{T}}{\alpha}=0\,,
\end{equation}
which makes $\lambda_{T}$ vanish. Then in this case, we find that
the stress energy tensor of VCDM, given in Eq.\ \eqref{eq:T_mu_nu_VCDM}
of Appendix \ref{sec:Covariant-VCDM-Tmunu}, can be rewritten as 
\begin{equation}
{\mathcal{T}_{{\bf v}}}_{~\nu}^{\mu}=-\frac{1}{4}\,\Mpl^{2}\,[4V(\phi_{0})+\lambda_{0}(3\lambda_{0}+4\phi_{0})]\,\delta^{\mu}{}_{\nu}=-\Mpl^{2}\Lambda\,\delta^{\mu}{}_{\nu}\,,
\end{equation}
where the effective cosmological constant on this background is given
by 
\begin{equation}
\Lambda=\frac{3}{4}\,\lambda_{0}^{2}+\lambda_{0}\phi_{0}+V(\phi_{0})\,,
\end{equation}
which agrees with Eq.\ (\ref{eq:phi_K_const_Lambda}).

In summary this shows that all GR solutions, written in the constant-$K$
slicing (whenever this choice of slicing is allowed), are also solutions
of VCDM. An example of this case is given in \cite{DeFelice:2021xps},
where the extrinsic curvature of the solutions is vanishing, finding
indeed that the static profile of spherically symmetric stars solutions
are also solutions of VCDM.

Motivated from sub-section \ref{sub-sec:weak_filed}, a more general
PPN treatment, which holds at higher order in the post Newtonian expansion,
can be performed by looking at the effective $\mathcal{T}_{{\bf v}}{}^{\mu}{}_{\nu}$
of VCDM found by using the covariant Lagrangian of the previous section,
and which is written in Eq.\ (\ref{eq:T_mu_nu_VCDM}) of appendix
\ref{sec:Covariant-VCDM-Tmunu}.

\subsection{Cosmological solutions}

Here we look at the dynamics of the cosmological background endowed
with Friedmann-Lemaître-Robertson-Walker (FLRW) metric and nonzero
spatial curvature. The three dimensional spatial metric is given as
\begin{equation}
\text{d}s_{3}^{2}=\left[a^{2}\,\frac{\text{d}r^{2}}{1-\kappa r^{2}}+\Phi\right]\text{d}r^{2}+a^{2}r^{2}(1+\zeta)\left\{ \frac{\text{d}z^{2}}{(1-z^{2})}+(1-z^{2})\,\text{d}\theta_{2}{}^{2}\right\} \,,\label{eqn:perturbedFLRW3dmetric}
\end{equation}
where $\kappa$ is the curvature constant and the terms in the curly
bracket define the two dimensional line element of a unit-radius sphere,
being $z=\cos\theta_{1}$, namely 
\begin{equation}
\text{d}\Omega^{2}\equiv\text{d}\theta_{1}{}^{2}+\sin^{2}\theta_{1}\,\text{d}\theta_{2}{}^{2}=\frac{\text{d}z^{2}}{(1-z^{2})}+(1-z^{2})\,\text{d}\theta_{2}{}^{2}\,,
\end{equation}
and the lapse is instead defined as 
\begin{equation}
N=\bigl[-^{(4)}g^{00}\bigr]^{-1/2}=\bar{N}(t)\,(1+\alpha)\,,
\end{equation}
whereas the shift contributes, on a homogeneous and isotropic background,
only perturbatively as follows 
\begin{equation}
N^{i}\partial_{i}=N^{r}\partial_{r}\equiv\chi\,\partial_{r}\,.\label{eqn:perturbedFLRWshift}
\end{equation}
The field variables $\alpha,\,\chi,\,\Phi,\,\zeta$, are linear perturbations
and have been introduced in order to derive the background equations
of motion.

We also define the scalar and vector fields in the VCDM Lagrangian
as in 
\begin{equation}
\phi=\bar{\phi}(t)+\delta\phi\,,\qquad\lambda=\bar{\lambda}(t)+\delta\lambda\,,\qquad\lambda_{{\rm gf}}^{i}\partial_{i}=[\bar{\lambda}_{2}(t,r)+\delta\lambda_{2}]\partial_{r}\,,
\end{equation}
where a bar stands for background quantities.

Now we include the matter fields using a Schutz-Sorkin Lagrangian~\citep{Schutz:1977df,Pookkillath:2019nkn}
for each matter component, namely

\begin{equation}
S_{I}=-\int\text{d}^{4}xN\sqrt{\gamma}\left[\rho_{I}(n_{I})+J_{I}\partial_{t}l_{I}+J_{I}^{i}\partial_{i}l_{I}\right]\,,
\end{equation}
where we have named, for each matter component labeled by $I$, the
0-th component of the vector field $J_{I}^{\mu}$ as $J_{I}^{0}=J_{I}$.
In the above Lagrangian, $\rho_{I}$ is energy density of each matter
component, $J_{I}^{\mu}$ is conserved number current density of the
matter, i.e.\ $J_{I}^{\mu}=n_{I}u_{I}^{\mu}$ and $\nabla_{\mu}J_{I}^{\mu}=0$,
whereas $l_{I}$ is the field variable related to the scalar part
of the velocity of the matter component. We have also defined the
number density of the fluid as 
\begin{equation}
n_{I}\equiv\sqrt{-g_{\mu\nu}J_{I}^{\mu}J_{I}^{\nu}}=\sqrt{-\left[(N^{i}N_{i}-N^{2})J_{I}^{2}+2N_{i}J_{I}^{i}J_{I}+\gamma_{ij}J_{I}^{i}J_{I}^{j}\right]}\,.
\end{equation}
We find it useful to introduce on the FLRW background a decomposition
for the matter fields given as follows 
\begin{align}
J_{I} & =\bar{J}_{I}(t)/\bar{N}(t)+\delta J_{I}\,,\qquad l_{I}\equiv\bar{l}_{I}(t)+\delta l{}_{I}\,,\qquad J_{I}^{i}\partial_{i}=\delta J_{I}^{r}\partial_{r}\,.
\end{align}
Here we have imposed homogeneity and isotropy in order to set the
three dimensional fluid velocity to vanish, i.e.\ $\bar{u}^{i}=0$,
or $\bar{J}^{i}=\bar{u}^{i}/\bar{n}=0$, where $n$ the fluid number
density on the background is only a function of time, i.e.\ $n=\bar{n}(t)$
on the background. This leads to also $\rho_{I}=\rho_{I}(n_{I})$
and $P_{I}=P_{I}[\rho_{I}(n_{I})]$ to be only functions of time on
the background, where $P_{I}$ is pressure of the matter component.

For the matter sector we have the following background equations of
motion 
\begin{equation}
\bar{n}_{I}=\bar{J}_{I}=\frac{N_{I,{\rm tot}}}{a^{3}}\,,\qquad\frac{\dot{\rho}_{I}}{\bar{N}}+3\,H\,(\rho_{I}+P_{I})=0\,,\qquad\bar{l}_{I}=-\int_{0}^{t}\bar{N}(t')\,\rho_{I,n_{I}}[\bar{n}(t')]\text{d}t'\,,
\end{equation}
where $N_{I.{\rm tot}}$ is a constant, corresponding to the constant
number of $I$-fluid particles for each matter component, whereas
$P_{I}=n_{I}\rho_{I,n_{I}}-\rho_{I}$ is the pressure of the $I$-fluid
component, and, finally, $H=\dot{a}/(aN)$ is the Hubble parameter.
For the gravity sector we have instead the following equations of
motion 
\begin{eqnarray}
E_{1} & \equiv & -\phi^{2}+3V(\phi)+\frac{3\rho}{\Mpl^{2}}=0\,,\label{eq:FRD_vcdm}\\
E_{2} & \equiv & -\frac{\dot{\phi}}{\bar{N}}+\frac{3}{2\Mpl^{2}}\,(\rho+P)=0\,,\label{eq:2ndEST_vcdm}
\end{eqnarray}
and the total conservation equation 
\begin{equation}
E_{3}\equiv\frac{\dot{\rho}}{\bar{N}}+3\,H\,(\rho+P)=0\,,
\end{equation}
where we have defined $\rho$ as the total effective energy density,
namely 
\begin{eqnarray}
\rho & \equiv & \sum_{I}\rho_{I}+\rho_{\kappa}\,,\label{eqn:def-rho}\\
\rho_{\kappa} & \equiv & -\frac{3\Mpl^{2}\kappa}{a^{2}}\,,\\
P & \equiv & \sum_{I}P_{I}+P_{\kappa}\,,\\
P_{\kappa} & \equiv & \frac{\Mpl^{2}\kappa}{a^{2}}\,,\label{eqn:def-Pkappa}
\end{eqnarray}
which also implies that $\dot{\rho}_{\kappa}/\bar{N}+3H\,(\rho_{\kappa}+P_{\kappa})=0$.
Note that for these background equations, the role of the curvature
amounts to giving an extra effective component for the term $\rho+P$.
This implies that, even in the absence of standard matter fields components,
we still have a non trivial dynamics for $\phi(t)$ in non-flat FLRW
solutions.

Now we also have for the background that 
\begin{equation}
K_{ij}=H\,\gamma_{ij}\,,\qquad{\rm and}\qquad K=3H\,,
\end{equation}
which, together with the equation of motion for $\lambda$ gives 
\begin{equation}
\frac{2}{3}\phi+2H+\lambda=0\,.
\end{equation}
By considering a combination of $E_{1}$, $E_{2}$, and $E_{3}$ we
obtain the following equation, 
\begin{equation}
\left(3H-\frac{3}{2}V_{,\phi}+\phi\right)(\rho+P)=0\,.
\end{equation}
Since in general $\rho+P\neq0$, this leads to 
\begin{equation}
3H+\phi=\frac{3}{2}V_{,\phi}\,.\label{eq:Bianchi_vcdm}
\end{equation}
The last equation of motion, the one for $\delta\phi$, is 
\begin{equation}
D_{i}\lambda_{{\rm gf}}^{i}=0\,,
\end{equation}
which implies that $\lambda_{{\rm gf}}^{i}$ vanishes, otherwise $\bar{\lambda}_{2}$
would be singular at $r=0$. The fact that in general both $\dot{\phi}$
and $K$ do not vanish leads to the consequence that on the cosmological
background VCDM solutions are different from $\Lambda$CDM, except
for the special case of a linear potential\footnote{Instead a quadratic potential, namely $V(\phi)=\beta_{0}+\beta_{1}\phi+\tfrac{1}{2}\,\beta_{2}\,\phi^{2}$
would instead lead, for $\beta_{2}<2/3$, to a $\Lambda$CDM background
with an effective redefined cosmological-background-Planck mass $M_{v}^{2}=2\Mpl^{2}/(2-3\beta_{2})$,
but still with $G_{{\rm eff}}=G_{N}$ for dust perturbations.} $V(\phi)=\beta_{0}+\beta_{1}\phi$.

We now prove that for any given/desired dynamics $H(z)$, with $H>0$,
the VCDM potential is in general always re-constructable, even in
the presence of a nonzero spatial curvature in the 3-D metric, generalizing
the result previously found in \citep{DeFelice:2020eju}. Let us rewrite
then the second Friedmann equation and the matter equation of motion
with the e-fold number $\mathcal{N}\equiv\ln(a/a_{0})$, by assuming
a known matter sector and on imposing a given dynamics for the Hubble
factor, i.e.\ $H=H(\mathcal{N})$. 
\begin{equation}
\frac{\text{d}\phi}{\text{d}\mathcal{N}}=\frac{3}{2}\frac{\rho+P}{H\Mpl^{2}}\,,\label{eq:2ndFried_N}
\end{equation}

Integrating Eq.~(\ref{eq:2ndFried_N}) with respect to $\mathcal{N}$
we get 
\begin{equation}
\phi(\mathcal{N})=\phi_{0}+\frac{3}{2}\frac{1}{\Mpl^{2}}\int_{0}^{\mathcal{N}}\frac{\rho(\mathcal{N}')+P(\mathcal{N}')}{H(\mathcal{N}')}\,\text{d}\mathcal{N}'\,.
\end{equation}
Now, on assuming that 
\begin{equation}
\rho+P>0\,,\qquad H>0\,,\label{eqn:rho+P}
\end{equation}
i.e.\ $H$ is positive definite as well as the total matter-curvature
contribution for $\rho+P$, the found function $\phi(\mathcal{N})$
is an increasing function of $\mathcal{N}$. Hence, there exists a
unique inverse function 
\begin{equation}
\mathcal{N}=\mathcal{N}(\phi)\,.
\end{equation}
Then, on using the first Friedmann equation Eq.~(\ref{eq:FRD_vcdm}),
we can finally write 
\begin{equation}
V=\frac{\phi^{2}}{2}+\frac{\rho\bigl(\mathcal{N}(\phi)\bigr)}{\Mpl^{2}}\,.
\end{equation}
Notice that the potential is not uniquely defined, since there is
a free choice for the constant $\phi_{0}$. In the spatially flat
case we have to impose the null energy condition as already mentioned
in~\citep{DeFelice:2020eju}. On the other hand, in the spatially
curved case $\rho$ and $P$ include contributions from the curvature
term (see (\ref{eqn:def-rho})-(\ref{eqn:def-Pkappa})) and thus (\ref{eqn:rho+P})
is either stronger or weaker than the null energy condition, depending
on the sign of the spatial curvature.

We also discuss the existence of the McVittie solution in VCDM theory
in the appendix \ref{sec:McVittie-solution}.


\section{Comparison between VCDM \& Cuscuton\label{sec:vcdm_and_cuscuton}}


VCDM theory and Cuscuton theory are sharing similar properties: in
both theories there are no additional degrees of freedom other than
that of GR, so that it is natural to ask if the solution of theses
theories share the same solutions or not. In a more mathematical language,
we ask if there exist a well-defined mapping from solutions of VCDM
to Cuscuton theory and vice versa.

At first we discuss the Cuscuton theory itself. The covariant action
for the Cuscuton theory is given by 
\begin{equation}
S=\int\text{d}^{4}x\sqrt{-g}\left[\frac{\Mpl^{2}}{2}{}^{(4)}R+\mu^{2}\sqrt{-X}-U(\varphi)\right]+S_{{\rm m}}\,,\label{eq:cus_action}
\end{equation}
where $S_{{\rm m}}$ represents the contribution from standard matter
fields, and we consider only the case of a timelike field $\varphi$,
so that $X<0$\footnote{In fact, we can consider an opposite sign convention, but here we
follow the $(-,+,+,+)$ convention for the metric and demand the Cuscuton
field $\varphi$ to be timelike.}, where 
\begin{equation}
X\equiv g^{\mu\nu}\partial_{\mu}\varphi\partial_{\nu}\varphi\,.
\end{equation}
Now, from the above Cuscuton action, we have the covariant equations
of motion 
\begin{align}
\Mpl^{2}G_{\mu\nu} & =T_{\mu\nu}-g_{\mu\nu}U+\frac{\mu^{2}}{\sqrt{-X}}\left[\partial_{\mu}\varphi\partial_{\nu}\varphi-g_{\mu\nu}X\right]\,,\\
U_{,\varphi} & =\frac{\mu^{2}}{\sqrt{-X}}\left[g^{\mu\nu}\nabla_{\mu}\nabla_{\nu}\varphi-\frac{1}{2X}\nabla_{\mu}X\nabla^{\mu}\varphi\right]\,,
\end{align}
where $T_{\mu\nu}$ represents the total stress energy tensor for
the matter fields, which satisfies the usual conservation equations
$\nabla_{\mu}T^{\mu}{}_{\nu}=0$.


\subsection{Cuscuton cosmology with quadratic potential\label{subsec:U_quadratic_cuscuton}}


Considering an homogeneous and isotropic FLRW metric with nonzero
spatial curvature we have the following equations of motion 
\begin{align}
U_{,\varphi} & =-3\mu^{2}H\text{sign}(\dot{\varphi})\,,\label{eq:EOM-cus}\\
H^{2} & =\frac{\rho_{\text{m}}+U}{3\Mpl^{2}}-\frac{\kappa}{a^{2}}\,,\label{eq:1stFriedmann-cus}\\
\dot{H} & =-\frac{\rho_{\text{m}}+P_{\text{m}}}{2\Mpl^{2}}-\frac{\mu^{2}|\dot{\varphi}|}{2\Mpl^{2}}+\frac{\kappa}{a^{2}}\,.\label{eq:2stFriedmann-cus}
\end{align}
Here we will assume that $\dot{\varphi}\neq0$, so that $\dot{\varphi}$
does not change its sign during the evolution of the universe. However,
we will also discuss the limiting case, namely $\dot{\varphi}\rightarrow0$,
and determine the conditions for which this limit can be taken while
the theory remains a valid effective field theory.

Using Eqs.\ ($\ref{eq:EOM-cus}$) and ($\ref{eq:1stFriedmann-cus}$),
we obtain Eq.\ ($\ref{eq:2stFriedmann-cus}$), after assuming the
standard energy conservation in the matter sector, namely $\dot{\rho}_{{\rm m}}+3H(\rho_{{\rm m}}+P_{{\rm m}})=0$.
We then have to solve only two independent equations, for instance
Eqs.\ ($\ref{eq:EOM-cus}$) and ($\ref{eq:1stFriedmann-cus}$). Indeed,
from Eqs.\ ($\ref{eq:EOM-cus}$) and ($\ref{eq:1stFriedmann-cus}$),
we find that the following equation always holds 
\begin{align}
U-\frac{\Mpl^{2}U_{,\varphi}^{2}}{3\mu^{4}} & =\frac{3\Mpl^{2}\kappa}{a^{2}}-\rho_{{\rm m}}\,.\label{eq:constr_cusc}
\end{align}

On assuming the following form for the potential 
\begin{equation}
U=U_{0}+\frac{1}{2}m^{2}\varphi^{2}\,,\label{eqn:quadraticpot_cuscuton}
\end{equation}
we find that Eq.\ (\ref{eq:constr_cusc}) leads to 
\begin{equation}
\frac{1}{2}\left[1-\frac{2\Mpl^{2}m^{2}}{3\mu^{4}}\right]m^{2}\varphi^{2}=\frac{3\Mpl^{2}\kappa}{a^{2}}-\rho_{{\rm m}}-U_{0}\,.
\end{equation}
(We shall study cosmology with a general potential in subsection \ref{subsec:cosmology}.)
Using this equation for $\varphi$, we rewrite the Friedmann equation
($\ref{eq:1stFriedmann-cus}$) as

\begin{equation}
3M_{c}^{2}H^{2}=\rho_{{\rm m}}+U_{0}-\frac{3\Mpl^{2}\kappa}{a^{2}}\,,\label{eq:cusco_phi2}
\end{equation}
where 
\[
M_{c}^{2}\equiv\Mpl^{2}-\frac{3\mu^{4}}{2m^{2}}\,,
\]
provided that 
\begin{equation}
m^{2}<0\,,\qquad{\rm or}\qquad m^{2}>\frac{3}{2}\,\frac{\mu^{4}}{\Mpl^{2}}\,.
\end{equation}
Notice that we have found an equation of motion, Eq.\
(\ref{eq:cusco_phi2}), which on the background, up to a redefinition
of the background effective gravitational constant, is identical to
the Friedmann equation in $\Lambda$CDM. However, it can be shown
that the growth of structure for this theory will still feel the standard
Newtonian gravitational constant, $G_{N}$. Hence, both the background
and the perturbations overall differ from $\Lambda$CDM. We can further
perform a time redefinition as $t=(M_{c}/\Mpl)\,\tilde{t}$, as to
make the Friedmann equation take the same form as in GR, namely 
\begin{equation}
3\Mpl^{2}\left(\frac{1}{a}\,\frac{da}{d\tilde{t}}\right)^{2}=\rho_{{\rm m}}+U_{0}-\frac{3\Mpl^{2}\kappa}{a^{2}}\,,
\end{equation}
out of which one can deduce the known GR solutions in terms of $a(\tilde{t})$.
For instance, in vacuum, on calling $U_{0}\equiv3\Mpl^{2}\tilde{H}_{0}^{2}$,
we find 
\begin{equation}
a(\tilde{t})\propto\begin{cases}
\cosh[\tilde{H}_{0}\tilde{t}] & {\rm for}\qquad\kappa=1\,,\\
\exp[\tilde{H}_{0}\tilde{t}] & {\rm for}\qquad\kappa=0\,,\\
\sinh[\tilde{H}_{0}\tilde{t}] & {\rm for}\qquad\kappa=-1\,,
\end{cases}
\end{equation}
as expected\footnote{In the case of $U_{0}=0$ and $\rho_{m}=0$ we the find for $\kappa=-1$
a Milne-like universe ($a=\tilde{t}$), which differs from the GR's
one ($a=t$) because of the different time rescaling.}. Here the $\kappa=0$ solution should be discarded, as leading to
a constant $\varphi$. However, if $\kappa=\pm1$, then $H$ becomes
time dependent, as well as $\varphi$, and these solutions can be
accepted for the Cuscuton theory.

Let us recast the effective Friedmann equation, Eq.\ (\ref{eq:cusco_phi2}),
in another way which is more suitable for phenomenology. Indeed let
us write 
\begin{equation}
3\Mpl^{2}H^{2}=\rho_{m}+\rho_{\Lambda}-\frac{3\Mpl^{2}\kappa}{a^{2}}+\rho_{{\rm cusc}}\,,
\end{equation}
where 
\begin{eqnarray}
\rho_{{\rm cusc}} & \equiv & 3(\Mpl^{2}-M_{c}^{2})\,H^{2}\,,\\
\rho_{\Lambda} & \equiv & U_{0}\,.
\end{eqnarray}
Then we have that 
\begin{equation}
1=\Omega_{m}+\Omega_{\Lambda}+\Omega_{\kappa}+\Omega_{{\rm cusc}}\,,
\end{equation}
where $\Omega_{m}=\rho_{m}/(3\Mpl^{2}H^{2})$, $\Omega_{\Lambda}=\rho_{\Lambda}/(3\Mpl^{2}H^{2})$,
$\Omega_{\kappa}=-\kappa/(a^{2}H^{2})$, and $\Omega_{{\rm cusc}}=1-M_{c}^{2}/\Mpl^{2}$.
This shows that $\Omega_{{\rm cusc}}={\rm constant}$, which will
prevent in general the other components' $\Omega$ to become unity
when they dominate the dynamics. The parameter $\Omega_{{\rm cusc}}$
corresponds to an additional free parameter of the Cuscuton theory
(with a quadratic potential), on which one can set in general constraints.


\subsection{Unacceptable solutions of Cuscuton theory}


In this section we discuss the $\Lambda$GR solutions which are not
acceptable solutions of Cuscuton theory, but, as previously shown,
acceptable in the VCDM theory.

\subsubsection{Static spherically symmetric solutions of VCDM}

Here we consider, for simplicity, spherically symmetric static solutions
of VCDM found in~\cite{DeFelice:2020onz}. The Cuscuton theory does
not allow for such a solutions, even outside the unitary gauge choice,
as $\varphi$ must be timelike and the presence of the potential does
not allow staticity for the spherically symmetric solutions of the
theory. In particular for such existing VCDM solutions we have $\phi={\rm constant}$,
and 
\begin{equation}
D_{i}\lambda_{{\rm gf}}^{i}=\frac{3V_{,\phi}+6b_{0}-2\phi}{F(r)}\neq0\,,\label{eqn:Dlambdagf}
\end{equation}
where $b_{0}=-K/3$, $K$ being the extrinsic curvature and, $F(r)$
is the $rr$ component of the spherically symmetric static metric.
The expression (\ref{eqn:Dlambdagf}) does not vanish in general.
Therefore these solutions have constant $K$ and $\phi$ but in general
$D_{i}\lambda_{{\rm gf}}^{i}\neq0$. As shown in \cite{Aoki:2021zuy},
all Cuscuton solutions are also solutions of VCDM provided that $D_{i}\lambda_{{\rm gf}}^{i}=0$
(as well as imposing that $V_{\phi\phi}\neq0$ while $\varphi$ remains
timelike). Here, in addition to the fact that $\phi$ is constant,
these solutions have in general a non-vanishing $D_{i}\lambda_{{\rm gf}}^{i}$,
which makes them outside the reach of Cuscuton theory. Nonetheless,
these solutions still belong to the class where VCDM admits $\Lambda$GR
solutions (because both $\phi$ and $K$ are constant in time and
space). Indeed, the static solutions found in \cite{DeFelice:2020onz}
are nothing but the Schwarzschild-de Sitter solutions only written
in a $K$-constant slicing coordinate system. The time-dependent spherically
symmetric solutions found in \cite{DeFelice:2020onz} have both $\dot{K}\neq0$
and $D_{i}\lambda_{{\rm gf}}^{i}\neq0$, so that they represent intrinsic-VCDM
solutions, i.e.\ solutions which are outside both GR and the Cuscuton
theory.


\subsubsection{GR vacuum solutions}


Let us consider now $\Lambda$GR vacuum solutions, that is four dimensional
solutions for the metric $g_{\mu\nu}$ which satisfy the following
tensorial equations of motion 
\begin{equation}
G_{\mu\nu}=-\Lambda\,g_{\mu\nu}\,,
\end{equation}
and we seek the condition for these solutions to hold also in the
Cuscuton theory. Before we look into the answer of this problem, let
us rewrite the Cuscuton action as proposed in~\citep{Bhattacharyya:2016mah},
namely 
\begin{equation}
S_{g}=\int\text{d}^{4}x\sqrt{-g}\left[\frac{\Mpl^{2}}{2}{}^{(4)}R+\mu^{2}u^{\mu}\nabla_{\mu}\varphi-U(\varphi)+\frac{\Mpl^{2}}{2}\,\sigma\,(g_{\mu\nu}u^{\mu}u^{\nu}+1)\right],\label{cus_action_sigma}
\end{equation}
out of which we can find covariant equations of motion for the metric
as 
\begin{equation}
G_{~\nu}^{\mu}=\frac{1}{\Mpl^{2}}\,{\mathcal{T}_{{\bf c}}}_{~\nu}^{\mu}\,,
\end{equation}
where 
\begin{eqnarray}
{\mathcal{T}_{{\bf c}}}_{~\nu}^{\mu} & = & [\mu^{2}u^{\alpha}\nabla_{\alpha}\varphi-U(\varphi)]\,\delta_{~\nu}^{\mu}+\Mpl^{2}\sigma u^{\mu}u_{\nu}\,,\\
\mu^{2}\nabla_{\mu}\varphi+\Mpl^{2}\sigma u_{\mu} & = & 0\,,\label{eq:cusc_u_eq}\\
g_{\mu\nu}u^{\mu}u^{\nu} & = & -1\,.\label{eq:uu-1}
\end{eqnarray}
On using $u_{\mu}=-\mu^{2}/(\Mpl^{2}\sigma)\,\nabla_{\mu}\varphi$,
and multiplying Eq.\ (\ref{eq:cusc_u_eq}) by $u^{\mu}$ we find
\begin{equation}
-g_{\mu\nu}\,\nabla^{\mu}\varphi\nabla^{\nu}\varphi=\frac{\Mpl^{4}}{\mu^{4}}\,\sigma^{2}\,,
\end{equation}
or 
\begin{equation}
\sigma=\frac{\mu^{2}}{\Mpl^{2}}\,\sqrt{-X}\,,
\end{equation}
where we have chosen the positive square root for $\sigma$. Then
in this case $u_{\mu}=-\frac{\nabla_{\mu}\varphi}{\sqrt{-X}}$, as
expected. In this case, for $\Lambda$GR vacuum solutions which are
also solution for the Cuscuton theory, we need to set 
\begin{equation}
{\mathcal{T}_{{\bf c}}}_{~\nu}^{\mu}=-\Mpl^{2}\Lambda\,\delta_{~\nu}^{\mu}\,,
\end{equation}
which leads to imposing 
\begin{equation}
Z_{\mu\nu}\equiv[\mu^{2}u^{\alpha}\nabla_{\alpha}\varphi+\Mpl^{2}\Lambda-U(\varphi)]\,g_{\mu\nu}+\Mpl^{2}\sigma u_{\mu}u_{\nu}=0\,.
\end{equation}
Then $g^{\mu\nu}Z_{\mu\nu}=0$ and $Z_{\mu\nu}u^{\mu}u^{\nu}=0$,
upon using (\ref{eq:uu-1}), imply that 
\begin{eqnarray}
4[\mu^{2}u^{\alpha}\nabla_{\alpha}\varphi+\Mpl^{2}\Lambda-U(\varphi)]-\Mpl^{2}\sigma & = & 0\,,\\
-[\mu^{2}u^{\alpha}\nabla_{\alpha}\varphi+\Mpl^{2}\Lambda-U(\varphi)]+\Mpl^{2}\sigma & = & 0\,,
\end{eqnarray}
which lead in particular to 
\begin{equation}
3\Mpl^{2}\sigma=0\,,\qquad X=0\,.
\end{equation}
This clearly contradicts the basic requirement of timelike $\partial_{\mu}\varphi$,
and thus cannot be accepted in the Cuscuton theory. So these $\Lambda$GR
solutions do not exist in Cuscuton. In particular, this result excludes
exact Minkowski, de Sitter or Schwarzschild-de Sitter solutions, as
the solution $X=0$ cannot be accepted. The same solution would be
instead accepted for quintessence models for which the configuration
$X=0$ is allowed.



\subsubsection{GR solutions in the presence of matter fields}


Let us consider also exact GR, in the presence of matter fields, that
is solutions of the following Einstein equations 
\begin{equation}
G_{~\nu}^{\mu}=-\Lambda\delta_{~\nu}^{\mu}+\frac{1}{\Mpl^{2}}\,T_{~\nu}^{\mu}\,,
\end{equation}
where $T_{\mu\nu}$ represent the total stress-energy tensor for matter
fields (which, by construction, we suppose to be minimally coupled
with gravity). On the other hand, a similar environment, in the Cuscuton
theory, would lead to the following equations of motion 
\begin{equation}
G_{~\nu}^{\mu}=\frac{1}{\Mpl^{2}}{\mathcal{T}_{{\bf c}}}_{~\nu}^{\mu}+\frac{1}{\Mpl^{2}}\,T_{~\nu}^{\mu}
\end{equation}
and once again we end up with the following necessary condition for
the $\Lambda$GR solutions to be solutions of the Cuscuton theory.
\begin{equation}
{\mathcal{T}_{{\bf c}}}_{~\nu}^{\mu}=-\Mpl^{2}\Lambda\,\delta_{~\nu}^{\mu}\,.
\end{equation}
Again this condition implies $X=0$, which is not acceptable for the
Cuscuton model. This results still holds even if in the Cuscuton theory
there is an explicit cosmological constant contribution $\Lambda_{c}$,
as this merely leads to a shift in the effective cosmological constant,
as in $\Lambda\to\Lambda-\Lambda_{c}$.


\subsubsection{Possible acceptable solutions close to GR solutions}


The Cuscuton field, by definition, is required to have timelike derivative.
This prevents the Cuscuton from admitting exact GR solutions. However,
it is possible in some situations that the field may be timelike but
may also be reaching an attractor for which $\dot{\varphi}\to0$.
Then we have a Cuscuton solution which is not exactly GR but very
close to it. In this case, it is necessary to understand whether or
not the Cuscuton theory still stands as a good effective low energy
theory. As to understand this point better we study the quantity $\delta X/X$
in linear perturbation theory in cosmology, adopting the ansatz (\ref{eqn:perturbedFLRW3dmetric})-(\ref{eqn:perturbedFLRWshift}),
and then determine which dynamics can give an acceptable behavior
for the perturbations fields. We achieve this goal by undoing the
unitary gauge, and using, instead the $\zeta=0$ gauge, which is always
well defined, as long as $H\neq0$. We also introduce a perfect fluid
as a matter field. Then we find, that on defining $\tilde{k}=k/(aH)$,
$w=P/\rho$, $\Omega=\rho/(3\Mpl^{2}H^{2})$, $c_{s}^{2}=\dot{P}/\dot{\rho}$,
we have after removing all the auxiliary fields that 
\begin{equation}
\frac{\delta X}{X}=\left[\frac{2\tilde{k}^{2}(1+3c_{s}^{2})}{2\tilde{k}^{2}+9(1+w)\Omega}-\frac{\ddot{\varphi}}{H\dot{\varphi}}\right]\frac{6\Omega}{2\tilde{k}^{2}+9(1+w)\Omega}\,\delta_{{\rm FG}}\,,
\end{equation}
where for simplicity we have fixed the background lapse function to
unity ($\bar{N}=1$), and have also assumed $\dot{\varphi}>0$. Here
we have also introduced the gauge invariant variable $\delta_{{\rm FG}}=\delta\rho/\rho-[\dot{\rho}/(H\rho)]\,\zeta$.
So in the limit $X\to0$, whether or not $\delta X/X$ blows up, hence
going out of the EFT validity, depends on the ratio $\frac{\ddot{\varphi}}{H\dot{\varphi}}$.
So even approaching $X=0$ does not necessarily mean that the theory
looses predictability. Indeed, we can choose dynamics, i.e.\ suitable
Cuscuton potentials, for which this ratio is always of order one,
leading to a consistent evolution of both the background and perturbations~\cite{Maeda:2022ozc}.
Otherwise, the EFT breaks down as the configuration approaches a GR
solution with or without matter fields.

It seems Cuscuton is doomed to be away from exact $\Lambda$CDM solutions,
but this does not necessarily mean that the theory is ruled out, as
we have already discussed above. Solutions might not be the same as
GR but close enough to them, in fact we could be even thinking of
cases for which, on the background, $T_{\mu\nu}^{c}\propto G_{\mu\nu}$,
giving a non-$\Lambda$CDM solution, which on the other hand could
be different from it only up to a redefinition of the effective Planck
mass for that particular background. This was indeed the case when
$U(\varphi)=U_{0}+\frac{1}{2}\,m^{2}\,\varphi^{2}$, as we have seen
in Sec.\ \ref{subsec:U_quadratic_cuscuton}. In this case though,
we should be seeing a difference between the cosmological effective
gravitational constant and the gravitational constant which determines
the evolution of dust perturbation, which is still $G_{N}$.


\subsection{Cosmology: VCDM vs Cuscuton}

\label{subsec:cosmology} 

As we have stated before, both VCDM and Cuscuton theories are MMG
Type-IIa theories, with only two propagating degrees of freedom, but
still both theories are different from GR, in general. Hence, it is
natural to check if the cosmology of these theories are related with
each other. In fact, since $D_{i}\lambda_{{\rm gf}}^{i}$ vanishes
and in general $\phi=\phi(t)$ with $\dot{\phi}\propto\rho+P\neq0$
(excluding an exact de Sitter case), one should expect to find a correspondence
between VCDM and the Cuscuton theory (see \citep{Aoki:2021zuy}).
Let us stress that this equivalence is accidental, and holds only
in particular cases, such as on a homogeneous and isotropic background.
As discussed so far, the two theories have different solutions and
as such the equivalence in general breaks.

In the following we will always consider both the conditions $H\neq0$
(standard cosmological background) and $\dot{\varphi}\neq0$ (always
holding at any finite time as to avoid EFT-breaking). Giving a FLRW
ansatz to the Cuscuton action Eq.\ (\ref{eq:cus_action}) we can
obtain the Cuscuton Friedmann equation 
\begin{equation}
H^{2}=\frac{1}{3\Mpl^{2}}\,[U(\varphi)+\rho]\,,
\end{equation}
on replacing $H$ by means of Eq.\ (\ref{eq:Bianchi_vcdm}), and
$\rho$, the total matter energy density, by means of the VCDM Friedmann
equation, namely Eq.\ (\ref{eq:FRD_vcdm}), we find

\begin{equation}
\left(\frac{1}{2}V_{,\phi}-\frac{1}{3}\phi\right)^{2}=\frac{1}{3\Mpl^{2}}\left[U(\varphi)+\frac{\Mpl^{2}}{3}(\phi^{2}-3V)\right].
\end{equation}
Imposing that this equation must hold at all times, we obtain 
\begin{eqnarray}
U(\varphi) & = & 3\Mpl^{2}\left(\frac{1}{2}V_{,\phi}-\frac{1}{3}\phi\right)^{2}-\frac{\Mpl^{2}}{3}\,(\phi^{2}-3V)\nonumber \\
 & = & \frac{3\Mpl^{2}}{4}V_{,\phi}^{2}+\Mpl^{2}\,(V-V_{,\phi}\phi)\,.\label{eq:trasf_VCDM_cusc}
\end{eqnarray}
So that 
\begin{eqnarray}
U_{,\varphi}\,d\varphi & = & \left[\frac{3}{2}V_{,\phi}-\phi\right]\Mpl^{2}V_{,\phi\phi}\,d\phi\nonumber \\
 & = & 3H\Mpl^{2}V_{,\phi\phi}\,d\phi\,.\label{eq:transform1}
\end{eqnarray}
On the other hand, the timelike Cuscuton satisfies also the following
condition 
\begin{equation}
U_{,\varphi}=-3\mu^{2}H\,{\rm sign}(\dot{\varphi})\,.
\end{equation}
Since now on we impose that during the known history of the universe
$H\neq0$, this implies that $U_{,\varphi}\neq0$. Then Eq.\ (\ref{eq:transform1})
becomes 
\begin{equation}
{\rm sign}(\dot{\varphi})\,d\varphi=-\frac{\Mpl^{2}}{\mu^{2}}\,V_{,\phi\phi}\,d\phi\,.\label{eq:dphic_dphiv}
\end{equation}
Therefore, we also require that $V_{,\phi\phi}\neq0$, for the mapping
to exist. As expected, this condition makes VCDM dynamics different
from $\Lambda$CDM.

Let us give an example for a well defined behavior of such a mapping.
Let us consider the case of a quadratic potential for the VCDM field,
namely 
\begin{equation}
V=\beta_{0}+\beta_{1}\phi+\frac{1}{2}\beta_{2}\phi^{2}\,.
\end{equation}
Then Eq.\ (\ref{eq:dphic_dphiv}) leads to 
\begin{equation}
{\rm sign}(\dot{\varphi})\,d\varphi=-\frac{\beta_{2}\Mpl^{2}}{\mu^{2}}\,d\phi\,,
\end{equation}
which can be integrated to give 
\begin{equation}
{\rm sign}(\dot{\varphi})\,\varphi=-\frac{\beta_{2}\Mpl^{2}}{\mu^{2}}\,\phi+\beta_{3}\,,
\end{equation}
and $\beta_{3}$ is a free constant of integration. Then on using
Eq.\ (\ref{eq:trasf_VCDM_cusc}), we find that on fixing the free
parameter $\beta_{3}$ as in 
\begin{equation}
\beta_{3}=\frac{\Mpl^{2}}{\mu^{2}}\,\frac{3\beta_{1}\beta_{2}}{2-3\beta_{2}}\,,
\end{equation}
the potential for the Cuscuton field can be written as 
\begin{equation}
U(\varphi)=U_{0}+\frac{1}{2}\,m^{2}\,\varphi^{2}\,,
\end{equation}
where 
\begin{eqnarray}
m^{2} & = & -\frac{\mu^{4}(2-3\beta_{2})}{2\Mpl^{2}\beta_{2}}\,,\\
\frac{U_{0}}{\Mpl^{2}} & = & \beta_{0}+\frac{3\beta_{1}^{2}}{4-6\beta_{2}}\,.
\end{eqnarray}
This Cuscuton potential agrees with the one in (\ref{eqn:quadraticpot_cuscuton})
and thus admits a $\Lambda$CDM background with an effective cosmological-gravitational-constant
$M_{c}^{2}$ which differs from $\Mpl^{2}$. Concretely, we have 
\begin{equation}
M_{c}^{2}=\Mpl^{2}-\frac{3\mu^{4}}{2m^{2}}=M_{v}^{2}\,,
\end{equation}
where $M_{v}^{2}=2\Mpl^{2}/(2-3\beta_{2})$. This is a working example
for which finding cosmological solutions in VCDM leads to knowing
mirror solutions in the Cuscuton theory and vice versa.

\section{summary and discussions\label{sec:Conclusions-and-discussion}}


In the present era, when some cosmological data seem to be either
inconsistent with each other or with General Relativity (GR), it is
of special interest to investigate the possibilities of modifying
gravity in several possible ways. In particular, since at solar system
scales no evidence has been found so far as to motivate the existence
of any new degree of freedom connected to the gravity sector, it makes
sense to look for those theories which do not add, by construction,
any new degree of freedom beside the two polarizations of gravitational
waves in the gravity sector. This possibility is now known to exist
in the framework of the so called ``minimally modified gravity''
(MMG). In particular, those theories which do not allow the existence
of an Einstein frame are called of Type II. Among these, we name of
Type-IIa those theories in which gravitational waves propagate, on
any background, at the speed of light.

Both VCDM and Cuscuton theories are Type-IIa MMG theories and we have
discussed here the relation between these two theories. In fact, both
theories on a cosmological background lead to an effective time dependent
extra energy-density component, which, however, does not lead to new
propagating degrees of freedom. This feature make them appealing as
to provide possibilities at solving e.g.\ the so called $H_{0}$-tension.

We have shown that the two theories in general are not equivalent.
We have in fact, explicitly shown this statement by mainly comparing
solutions which exist in VCDM but not in Cuscuton, as demonstrated
in Fig.~\ref{summaryFig}. The following two facts clearly show the
non-equivalence of the theories. First, the derivative of the Cuscuton
scalar needs to be always timelike on any background. As a consequence,
backgrounds which require the Cuscuton field $\varphi$ to be constant
(in time and space) are not acceptable solutions for this theory.
This situation takes place when we consider exact GR solutions (in
the presence of minimally coupled matter, including possibly a cosmological
constant). Although the Cuscuton field can lead to solutions which
are close to the GR counterparts, it does not allow for exact GR solutions
to be also solutions of the theory.

Second, on the other hand, we have shown that when a GR-solution (with
or without matter fields, in the presence of a cosmological constant)
allows for a foliation which is endowed with a constant trace of the
extrinsic curvature (both in time and space) then these same solutions
also are solutions in the VCDM theory. For instance, this result holds
true in VCDM for both the static Schwarzschild-de Sitter metric (for
any slicing admitting $K=K_{0}$) and the vacuum Kerr-de Sitter solutions
(in Boyer--Lindquist coordinates), since both solutions have a constant
trace for the extrinsic curvature $K$.

As a consequence, we also worked out various limits of the VCDM theory,
say weak field limit and the de Sitter limit of the VCDM theory and
show that all these limits are well defined, e.g.\ no strong coupling
is present, and exactly match the GR solutions.

We also find that in the context of cosmology these two theories are
always related in general to each other, since $D_{i}\lambda_{{\rm gf}}^{i}=\nabla^{2}\lambda_{2}=0$.
For a special form of potential, i.e., for a quadratic potential for
VCDM and Cuscuton, this mapping is well defined. However, the effective
Planck mass for cosmological backgrounds is modified to be $M_{v}^{2}=2\Mpl^{2}/(2-3\beta_{2})$.

In summary, we have confirmed that all acceptable Cuscuton solutions
are also solutions of VCDM (see e.g.\ \citep{Aoki:2021zuy}, and
in particular, cosmological solutions belong to this case). However,
in addition to these solutions, VCDM has other solutions which, as
mentioned above, are exact solutions of GR (with our without matter
fields and a cosmological constant) which are not, on the other hand,
acceptable solutions in Cuscuton. Finally, besides these, VCDM has
a third category of solutions, which consists of solutions which are
intrinsic only to VCDM, which are neither GR ($K$ is not a constant
in time or space) nor Cuscuton (no mapping in this case exists).

This study opens up several possible future directions. One direction
is to look for possible signatures coming from the properties of gravitational
waves propagating on intrinsic-VCDM background solutions. Another
direction worth investigating is, whether the VCDM theory can be recast
as an IR limit of some Lorentz breaking UV theory. It was shown in~\citep{Lacombe:2022cbq},
that using brane-world model with k-essence we can have a self-tuning
of the cosmological constant. Interestingly, the self-tuning mechanism
constraints the Lagrangian to spacelike Cuscuton. Hence, it is interesting
to explore the brane-world scenario with spacelike VCDM in five dimension
to see if a self-tuning mechanism is possible with VCDM theory.

\begin{acknowledgments}
The work of A.D.F.\ was supported by Japan Society for the Promotion
of Science Grants-in-Aid for Scientific Research No.\ 20K03969. K.M.
would like to acknowledges the Yukawa Institute for Theoretical Physics
at Kyoto University, where the present work was begun during the Visitors
Program of FY2021. The work of K.M. was supported by JSPS KAKENHI
Grant Numbers JP17H06359 and JP19K03857. The work of S.M. is supported
in part by Japan Society for the Promotion of Science Grants-in-Aid
for Scientific Research No. 17H02890. , and No. 17H06359 and by World
Premier International Research Center Initiative, The Ministry of
Education, Culture, Sports, Science and Technology, Japan. The work
of M.C.P.\ was supported by the Japan Society for the Promotion of
Science Grant-in-Aid for Scientific Research No.\ 17H06359. 
\end{acknowledgments}


\appendix


\section{Covariant VCDM equations of motion\label{sec:Covariant-VCDM-Tmunu}}

In this appendix, we explicitly write down all the equations of motion
for the VCDM covariant action introduced in Eq.\ \eqref{eq:cov_act},
and evaluate them, as an example, on a FLRW background. In the remaining
part of this section we find it convenient to perform the following
field redefinition 
\begin{equation}
\lambda_{T}\equiv\frac{\lambda_{N}}{\alpha^{2}}\,.
\end{equation}

The equation of motion for $\lambda_{T}$ (or, equivalently, for $\lambda_{N}$)
leads to 
\begin{equation}
\frac{E_{\lambda_{T}}}{\Mpl^{2}}\equiv1+\alpha^{2}\nabla_{\alpha}T\,\nabla^{\alpha}T=0\,,
\end{equation}
which, on a FLRW manifold on which $T=t$, gives 
\begin{equation}
\alpha(t)=N(t)\,,
\end{equation}
as expected (discarding the other solution $\alpha=-N$). The equation
of motion for $\lambda$ instead gives, on a general background, 
\begin{equation}
\frac{E_{\lambda}}{\Mpl^{2}}=\frac{3}{2}\,\lambda+\phi-\alpha\,\nabla_{\mu}\nabla^{\mu}T-\nabla_{\mu}\alpha\,\nabla^{\mu}T=0\,,
\end{equation}
which evaluated on FLRW returns 
\begin{equation}
\lambda=-\frac{2}{3}\,\phi-2\,H\,,\qquad{\rm with}\qquad H\equiv\frac{\dot{a}}{Na}\,.
\end{equation}
Next, let us consider the equation of motion for $\alpha$. It can
be written as 
\begin{equation}
\frac{E_{\alpha}}{\Mpl^{2}}=\frac{2\lambda_{N}}{\alpha^{3}}+\frac{\lambda_{2}}{\alpha^{2}}\,\nabla_{\mu}\nabla^{\mu}\phi+\nabla_{\mu}\lambda\,\nabla^{\mu}T+(\nabla_{\mu}\lambda_{2}\,\nabla^{\mu}T)(\nabla_{\nu}\phi\,\nabla^{\nu}T)+\lambda_{2}(\nabla^{\mu}T)(\nabla^{\nu}\phi)\nabla_{\mu}\nabla_{\nu}T=0\,,\label{eq:E_alpha}
\end{equation}
which can be used to fix $\lambda_{N}$ in terms of the other fields.
On doing this on FLRW we do find 
\begin{equation}
\lambda_{N}(t)=-\frac{1}{2}\,\frac{\dot{\phi}}{N}\,\dot{\lambda}_{2}+\frac{3\dot{a}\dot{\phi}+a\ddot{\phi}}{2aN}\,\lambda_{2}+\frac{1}{2}\,N\,\dot{\lambda}\,.
\end{equation}
On setting this constraint on $\lambda_{N}$, we have that the covariant
equation of motion for the $T$ field, 
\begin{eqnarray}
\frac{E_{T}}{\Mpl^{2}} & = & 2\lambda_{N}\nabla_{\nu}\nabla^{\nu}T-\alpha\nabla_{\nu}\nabla^{\nu}\lambda+2\nabla_{\mu}\lambda_{N}\,\nabla^{\mu}T-\nabla_{\mu}\lambda\,\nabla^{\mu}\alpha+\lambda_{2}\,(\nabla_{\mu}\phi\,\nabla^{\mu}\alpha)(\nabla_{\nu}\nabla^{\nu}T)\nonumber \\
 & + & \alpha\,(\nabla_{\mu}\phi\,\nabla^{\mu}\lambda_{2})(\nabla_{\nu}\nabla^{\nu}T)-\alpha\,(\nabla_{\mu}\phi\,\nabla^{\mu}T)(\nabla_{\nu}\nabla^{\nu}\lambda_{2})+\alpha\,\lambda_{2}\,(\nabla_{\nu}\nabla^{\nu}T)(\nabla_{\mu}\nabla^{\mu}\phi)\nonumber \\
 & + & \lambda_{2}\,(\nabla_{\mu}T\,\nabla^{\mu}\alpha)(\nabla_{\nu}\nabla^{\nu}\phi)+\alpha\,\lambda_{2}\,\nabla^{\mu}T\,(\nabla_{\nu}\nabla^{\nu}\nabla_{\mu}\phi)-(\nabla_{\mu}\lambda_{2}\,\nabla^{\mu}\alpha)(\nabla_{\nu}\phi\,\nabla^{\nu}T)\nonumber \\
 & + & \lambda_{2}\,(\nabla^{\mu}T)(\nabla^{\nu}\alpha)\nabla_{\mu}\nabla_{\nu}\phi+(\nabla_{\mu}\lambda_{2}\,\nabla^{\mu}\phi)(\nabla_{\nu}\alpha\,\nabla^{\nu}T)-\alpha\,\lambda_{2}\,R_{\mu\nu}\,(\nabla^{\mu}T)(\nabla^{\nu}\phi)\nonumber \\
 & - & 2\alpha\,(\nabla^{\mu}\lambda_{2})(\nabla^{\nu}\phi)\nabla_{\mu}\nabla_{\nu}T+\lambda_{2}\,(\nabla^{\mu}T)(\nabla^{\nu}\phi)\nabla_{\mu}\nabla_{\nu}\alpha=0\,,\label{eq:T-eom}
\end{eqnarray}
is automatically satisfied on FLRW. Let us now consider the equation
of motion for $\phi$. This can be written as 
\begin{equation}
\frac{E_{\lambda_{2}}}{\Mpl^{2}}=-\frac{1}{\alpha}\,\nabla_{\mu}\nabla^{\mu}\phi-\alpha\,(\nabla_{\mu}\phi\,\nabla^{\mu}T)\,(\nabla_{\nu}\nabla^{\nu}T)-(\nabla_{\mu}\alpha\,\nabla^{\mu}T)(\nabla_{\nu}\phi\,\nabla^{\nu}T)-\alpha\,(\nabla^{\mu}T)(\nabla^{\nu}T)(\nabla_{\mu}\nabla_{\nu}\phi)=0\,,
\end{equation}
which also identically vanishes on FLRW, as expected. We also need
to evaluate the equation of motion for $\lambda_{2}$, which reads
\begin{eqnarray}
\frac{E_{\phi}}{\Mpl^{2}} & = & \lambda+V_{,\phi}+\frac{\lambda_{2}}{\alpha^{2}}\,(\nabla_{\nu}\nabla^{\nu}\alpha)-\frac{1}{\alpha}\,(\nabla_{\nu}\nabla^{\nu}\lambda_{2})-\frac{2\lambda_{2}}{\alpha^{3}}\,(\nabla_{\mu}\alpha\,\nabla^{\mu}\alpha)+\frac{2}{\alpha^{2}}\,(\nabla_{\mu}\lambda_{2}\,\nabla^{\mu}\alpha)\nonumber \\
 & - & \alpha\,(\nabla_{\mu}T\,\nabla^{\mu}\lambda_{2})(\nabla_{\nu}\nabla^{\nu}T)-\alpha\,\lambda_{2}\,\nabla^{\mu}T\,(\nabla_{\nu}\nabla^{\nu}\nabla_{\mu}T)-(\nabla_{\mu}\lambda_{2}\,\nabla^{\mu}T)(\nabla_{\nu}\alpha\,\nabla^{\nu}T)\nonumber \\
 & - & \alpha\,(\nabla^{\mu}T)(\nabla^{\nu}T)\nabla_{\mu}\nabla_{\nu}\lambda_{2}-\lambda_{2}\,(\nabla^{\mu}\alpha)(\nabla^{\nu}T)\nabla_{\mu}\nabla_{\nu}T-2\alpha\,(\nabla^{\mu}\lambda_{2})(\nabla^{\nu}T)\nabla_{\mu}\nabla_{\nu}T\nonumber \\
 & - & \alpha\,\lambda_{2}\,(\nabla_{\mu}\nabla_{\nu}T)\,(\nabla^{\mu}\nabla^{\nu}T)=0\,,
\end{eqnarray}
and gives on the homogeneous background 
\begin{equation}
\phi-\frac{3}{2}\,V_{,\phi}+3H=0\,,
\end{equation}
but on another generic background it would be used as to fix $\lambda_{2}$.
Finally, let us evaluate the stress-energy tensor as 
\begin{eqnarray}
\frac{{\mathcal{T}_{{\bf v}}}_{~\nu}^{\mu}}{\Mpl^{2}} & = & \delta^{\mu}{}_{\nu}\{-V\,-\tfrac{3}{4}\,\lambda^{2}-\lambda\phi+\lambda_{2}\alpha^{-2}(\nabla_{\beta}\phi\,\nabla^{\beta}\alpha)-\alpha^{-1}(\nabla_{\beta}\phi\,\nabla^{\beta}\lambda_{2})\nonumber \\
 & - & \alpha(\nabla_{\beta}\lambda\,\nabla^{\beta}T)-\alpha(\nabla^{\gamma}\lambda_{2}\,\nabla_{\gamma}T)(\nabla_{\beta}\phi\,\nabla^{\beta}T)-\alpha\lambda_{2}\nabla^{\beta}\phi(\nabla_{\beta}\nabla_{\gamma}T)\nabla^{\gamma}T\}\nonumber \\
 & - & \nabla^{\mu}T\,\nabla_{\nu}T\,\{2\lambda_{N}+\alpha\lambda_{2}(\nabla_{\beta}\nabla^{\beta}\phi)+\lambda_{2}(\nabla_{\beta}\phi\,\nabla^{\beta}\alpha)+\alpha(\nabla_{\beta}\phi\,\nabla^{\beta}\lambda_{2})\}\nonumber \\
 & - & \frac{\lambda_{2}}{\alpha^{2}}(\nabla^{\mu}\phi\,\nabla_{\nu}\alpha+\nabla^{\mu}\alpha\,\nabla_{\nu}\phi)+\frac{1}{\alpha}(\nabla^{\mu}\phi\,\nabla_{\nu}\lambda_{2}+\nabla^{\mu}\lambda_{2}\,\nabla_{\nu}\phi)\nonumber \\
 & + & \alpha(\nabla^{\mu}T\,\nabla_{\nu}\lambda+\nabla^{\mu}\lambda\,\nabla_{\nu}T)+\alpha(\nabla^{\mu}T\,\nabla_{\nu}\phi+\nabla^{\mu}\phi\,\nabla_{\nu}T)\,(\nabla_{\beta}\lambda_{2}\,\nabla^{\beta}T)\nonumber \\
 & + & \alpha(\nabla^{\mu}T\,\nabla_{\nu}\lambda_{2}+\nabla^{\mu}\lambda_{2}\,\nabla_{\nu}T)\,(\nabla_{\beta}\phi\,\nabla^{\beta}T)\nonumber \\
 & + & \alpha\lambda_{2}\nabla^{\beta}T\,[\nabla^{\mu}\phi(\nabla_{\beta}\nabla_{\nu}T)+(\nabla_{\beta}\nabla^{\mu}T)\nabla_{\nu}\phi]\,.\label{eq:T_mu_nu_VCDM}
\end{eqnarray}
Then on constructing 
\begin{equation}
\Mpl^{2}G^{\mu}{}_{\nu}={\mathcal{T}_{{\bf v}}}_{~\nu}^{\mu}+\sum_{I}T_{(I)}^{\mu}{}_{\nu}\,,
\end{equation}
we find on a FLRW background 
\begin{eqnarray}
\frac{1}{3}\,\phi^{2}-V+\frac{3\kappa}{a^{2}} & = & \frac{1}{\Mpl^{2}}\sum_{I}\rho_{I}\,,\\
\frac{\dot{\phi}}{N}-\frac{1}{2}\,\phi^{2}+\frac{3}{2}\,V-\frac{3\kappa}{2a^{2}} & = & \frac{3}{2\Mpl^{2}}\sum_{I}P_{I}\,,
\end{eqnarray}
as expected.


\section{McVittie solution in VCDM\label{sec:McVittie-solution}}


In the following we consider the McVittie solution in VCDM. Let us
assume we have the following metric ansatz 
\begin{equation}
ds^{2}=-N(t)^{2}\left(\frac{2ar-m}{2ar+m}\right)^{2}dt^{2}+a^{2}\left(1+\frac{m}{2ar}\right)^{4}\left[dr^{2}+r^{2}\left(\frac{dz^{2}}{1-z^{2}}+(1-z^{2})\,d\theta_{2}{}^{2}\right)\right],
\end{equation}
with $m$ being a constant. We will also consider the matter content
only consists of a bare cosmological constant. In this case, by looking
for example at $R_{\mu\nu\alpha\beta}R^{\mu\nu\alpha\beta}$, we can
see the spacetime is not homogeneous. Therefore in this case, besides
the choice of coordinate 
\begin{equation}
T(t)=t\,,\qquad\textrm{as a choice of coordinates,}
\end{equation}
we suppose a spherically symmetric profile for all the fields in the
theory, namely 
\begin{equation}
\phi=\phi(t,r)\,,\qquad\alpha=\alpha(t,r)\,,\qquad\lambda=\lambda(t,r)\,,\qquad\lambda_{2}=\lambda_{2}(t,r)\,,\qquad\lambda_{T}=\lambda_{T}(t,r)\,.
\end{equation}
The equation of motion for $\lambda_{T}$ sets 
\begin{equation}
\alpha(t,r)=\frac{2ar-m}{2ar+m}\,N(t)\,,
\end{equation}
while, the equation of motion for $\lambda$ gives 
\begin{equation}
\lambda(t,r)=-\frac{2}{3}\,\phi(t,r)-2H\,,
\end{equation}
where we have defined 
\begin{equation}
H\equiv\frac{\dot{a}}{Na}\,,
\end{equation}
out of which the trace of the extrinsic curvature is given by $K=3H$.
Notice that at this level, we cannot impose homogeneity on $\phi$
or $\lambda$. Next solving the equation of motion for $\alpha$,
we find $\lambda_{T}$ as 
\begin{equation}
\lambda_{T}=\lambda_{T}(\partial_{t}\lambda_{2},\lambda_{2},\partial_{r}^{2}\phi,\partial_{t}\phi,\partial_{r}\phi,\dot{H},H,N,a,r)\,.
\end{equation}
The equation of motion for $\lambda_{2}$ sets the following constraint
\begin{equation}
(2ar+m)\,\partial_{r}^{2}\phi+4a\,\partial_{r}\phi=0\,,
\end{equation}
which can be solved for 
\begin{equation}
\phi(t,r)=\phi_{h}(t)+\frac{\phi_{n}(t)}{r+\frac{m}{2a}}\,,
\end{equation}
so $\phi$ in general might have an inhomogeneous contribution. In
principle, on matching the field $\phi$ with cosmological boundary
conditions would set $\lim_{r\to\infty}\phi(t,r)=\phi_{h}(t)$ giving
an homogeneous profile (so we cannot use the boundary conditions to
set $\phi_{n}(t)$ to vanish in this case). Therefore, we will keep
this solution as it is, and see whether the equations of motion set
the values of $\phi_{h}$ or $\phi_{n}$. In fact, since the Einstein
equations are 
\begin{equation}
\Mpl^{2}\,G^{\mu}{}_{\nu}=\mathcal{T}_{{\bf v}}{}^{\mu}{}_{\nu}\,,
\end{equation}
we find that the $(0,1)$ component of these equations lead to 
\begin{equation}
\frac{8\phi_{n}a^{2}}{N(3m^{2}-12a^{2}r^{2})}=0\,,
\end{equation}
which requires 
\begin{equation}
\phi_{n}(t)=0\,,
\end{equation}
leading to 
\begin{equation}
\phi(t,r)=\phi_{h}(t)=\phi(t)\,,
\end{equation}
or the field is homogeneous. At this level, looking at the $(0,0)$
component of the Einstein equations we find 
\begin{equation}
E_{1}\equiv\frac{1}{3}\,\phi(t)^{2}-V(\phi(t))=0\,.
\end{equation}
Therefore, for a generic potential \footnote{We will not consider here the possibility of a special form of a quadratic
potential such that $V=V_{0}+\frac{1}{3}\phi^{2}$.} we have 
\begin{equation}
\phi(t)=\phi_{0}={\rm constant}.
\end{equation}
All the non-diagonal components of the Einstein equations now vanish
whereas the $(i,i)$ components lead once again to the condition $E_{1}=0$.
Now all the Einstein equations are satisfied. At this level also the
equation of motion for $T$ is automatically satisfied. There is one
last equation of motion we need solve, the equation of motion for
$\phi$, which gives 
\begin{equation}
\partial_{r}^{2}\lambda_{2}+\frac{4a}{2ar+m}\,\partial_{r}\lambda_{2}=-N(t)\,\frac{(2ar-m)(2ar+m)^{3}}{48a^{2}r^{4}}\,[2\phi_{0}-3V_{,\phi}(\phi_{0})+6H(t)]\,.
\end{equation}
The equation can be solved as 
\begin{eqnarray}
\lambda_{2} & = & \lambda_{2,h}(t)+\frac{\lambda_{2,n}(t)}{(2ar+m)}+\frac{N\,[\frac{2}{3}\phi_{0}-3V_{,\phi}(\phi_{0})+2H(t)]}{96a^{2}r^{2}(2ar+m)}\,\bigl[-32a^{5}r^{5}-144ma^{4}r^{4}-96m^{2}a^{3}r^{3}\bigl(\ln r+\tfrac{2}{3}\bigr)\nonumber \\
 & - & 48m^{3}a^{2}r^{2}\bigl(\ln r+\tfrac{8}{3}\bigr)+18m^{4}ra+m^{5}\bigr]\,.
\end{eqnarray}
Notice that $\lambda_{2}$ grows as $r\to\infty$, although the source
and the Riemann tensor tends to vanish for large $r$'s. On fixing
boundary conditions so that $\lambda_{2}$ does not diverge at infinity
(as otherwise $\lambda_{2}\propto r^{2}$), we require 
\begin{equation}
\frac{2}{3}\phi_{0}-3V_{,\phi}(\phi_{0})+2H(t)=0\,,
\end{equation}
which states that 
\begin{equation}
H=H_{0}={\rm constant}.
\end{equation}
In this case the solution automatically reduces to the MacVittie's
solution obtained in GR, since $K$ becomes a constant. With these
boundary conditions, $\lambda_{2}$ reduces to the flat-de Sitter
solution, $\lambda_{2}=\lambda_{2}(t)$, in the limit $\frac{m}{ar}\ll1$.
Therefore the chosen boundary conditions for $\lambda_{2}$ make $\lambda_{2}$
match an homogeneous profile at infinity.

\bibliographystyle{apsrev4-2}
\bibliography{bibliography}

\end{document}